\documentclass[11pt]{article}
\usepackage{indentfirst}
\usepackage{float}

\usepackage{tikz}
\usetikzlibrary{shapes, fit, backgrounds, calc, positioning}
\usetikzlibrary{arrows.meta}
\usetikzlibrary{decorations.pathmorphing}

\usepackage[utf8]{inputenc}
\usepackage{amsmath}
\usepackage{amssymb}
\usepackage{amsthm}
\usepackage{cite}
\usepackage{braket}
\usepackage{booktabs}
\usepackage{arydshln}
\usepackage{xcolor}

\newtheorem{theorem}{Theorem}
\newtheorem{corollary}{Corollary}

\newtheorem{lemma}{Lemma}
\newtheorem{example}{Example}
\newtheorem{proposition}{Proposition}
\newtheorem{remark}{Remark}
\newenvironment{proofbody}{\par\noindent\ignorespaces}{\qed\par\medskip}

\newcommand\blfootnote[1]{%
  \begingroup
  \renewcommand\thefootnote{}\footnote{#1}%
  \addtocounter{footnote}{-1}%
  \endgroup
}

\newcommand{\Fq}{\mathbb{F}_q}
\newcommand{\Fp}{\mathbb{F}_p}
\newcommand{\F}{\mathbb{F}}
\newcommand{\ba}{\boldsymbol a}
\newcommand{\bb}{\boldsymbol b}
\newcommand{\bs}{\boldsymbol s}
\newcommand{\bw}{\boldsymbol w}
\newcommand{\by}{\boldsymbol y}
\newcommand{\bz}{\boldsymbol z}
\newcommand{\bxi}{\boldsymbol\xi}
\newcommand{\bx}{\boldsymbol x}
\newcommand{\cA}{\mathcal A}
\newcommand{\cB}{\mathcal B}
\newcommand{\cX}{\mathcal X}
\newcommand{\cR}{\mathcal R}
\DeclareMathOperator{\rank}{rank}
\DeclareMathOperator{\row}{row}
\DeclareMathOperator{\nullspace}{nullspace}

\title{Capacity Meets the Four Subspace Problem:\\ Distributed Linear Computation and the Cross-Ratio}
\author{Hua Sun}
\date{}

\begin{document}
\maketitle

\blfootnote{
Hua Sun (email: hua.sun@unt.edu) is with the Department of Electrical Engineering at the University of North Texas.}

\begin{abstract}
$K$ users each observe a fixed linear function of common data over a finite field of characteristic $p$, and a server must recover another fixed linear function with zero error from one message sent by each user. The capacity, the minimum total number of symbols the users send per computed instance, depends only on the configuration of subspaces formed by these linear functions. We study the instance in which three users observe $\ba$, $\bb$ and $\ba+\bb$ and the server wants $\ba+\lambda\bb$. The four functions span four lines in a plane with cross-ratio $\lambda$, the smallest instance of the four subspace problem of representation theory. It is the smallest configuration whose capacity is not determined by the dimensions of the subspaces formed by sums and intersections: these dimensions are the same for every $\lambda$, but the capacity depends on the cross-ratio and on the prime subfield. For the harmonic cross-ratios, $\lambda=-1$, $\lambda=2$ or $2\lambda=1$, it is $2$: every zero-error code reveals $(\ba,\bb)$ to the server, so no code beats having two users send their observations. For $\lambda$ outside the prime subfield it is the cut-set bound $3/2$, achieved by codes linear over the prime subfield, while codes linear over the field need $2$. For the other $\lambda$ in the prime subfield it lies in $[3p/(2p-1),2]$, so no code at any block length approaches the cut-set bound. The qualitative separation persists under vanishing error for uniform data: the cut-set bound is tight exactly for $\lambda$ outside the prime subfield. The four-line instance is also exact repair of one node of a $[4,2]$ MDS storage code, and the results translate to that setting.
\end{abstract}

\allowdisplaybreaks
\emergencystretch=3em
\section{Introduction}\label{sec:intro}

Linear computation is at the core of many distributed computing problems whose fundamental limits are studied through capacity. A server wants linear functions of data spread over users, workers or databases, and the problems differ in what they add to this linear core. Coded computing and gradient coding add stragglers: the server recovers a linear function, a sum of gradients or a matrix product from any large enough subset of workers~\cite{Lee_speeding,Tandon_gradient,Yu_polynomial,Wan_Sun_Ji_Caire_DLSC,Wan_Sun_Ji_Caire}. Secure aggregation and secure summation add security: the server learns the sum of the users' inputs and nothing more, possibly under dropouts and collusion~\cite{Bonawitz,Zhao_Sun_dropout,Zhao_Sun_summation}. Private computation and secure matrix multiplication add privacy of the demand or of the data~\cite{Sun_Jafar_private_computation,Chang_Tandon,Jia_Jafar_batch,Wan_Sun_Ji_Caire_secure}. Computation broadcast, introduced in~\cite{Sun_Jafar_CB}, and its linear version~\cite{Yao_Jafar_3LCBC,Yao_Jafar_generic} add side information at the receiving users, and cache-aided matrix retrieval adds caches~\cite{Wan_cache_matrix}. Removing these elements leaves a common problem: users observe linear functions of common data, and a server wants another linear function. This paper studies that problem by itself, and its capacity turns out to depend on more than the dimensions of the subspaces involved.

Consider $K$ users and one server (Figure~\ref{fig:model}(a)). The data are $\mu$ symbols of $\Fq$ per instance; user $k$ observes a fixed linear function $\mathbf H_k\bw$ of the data $\bw\in\Fq^\mu$, its \emph{observation}, and the server wants another fixed linear function $\mathbf D\bw$. Each user sends one message to the server over a noiseless link, and the server must recover $\mathbf D\bw$ with zero error; a code operates on $L$ independent instances, and $L$ is its \emph{block length}. The \emph{rate} of a user is the number of $q$-ary symbols it sends per instance, the \emph{sum-rate} is the total over the users, and the \emph{capacity} is the minimum achievable sum-rate. What makes the problem nontrivial is the overlap between the users' observations. If the $\mathbf H_k$ select disjoint coordinates of $\bw$, each user sends the part of $\mathbf D\bw$ that depends on its coordinates and the cut-set bound is tight; when the observations overlap, the users must decide who sends what the others also observe, and in what form. The overlaps are described by the subspaces spanned by the rows of the coefficient matrices, and the question is which features of these subspaces determine the capacity.

\begin{figure}[t]
\centering
\begin{tikzpicture}[
  user/.style={draw, rounded corners=2pt, minimum width=1.75cm, minimum height=0.62cm, font=\small},
  server/.style={draw, rounded corners=2pt, minimum width=1.25cm, minimum height=2.6cm, font=\small},
  msg/.style={-{Stealth[length=2mm]}, thick},
  lab/.style={font=\small, fill=white, inner sep=1pt}]
\begin{scope}
\node[user] (u1) at (0,1.05) {$\mathbf H_1\bw$};
\node[user] (u2) at (0,0.25) {$\mathbf H_2\bw$};
\node at (0,-0.38) {$\vdots$};
\node[user] (uK) at (0,-1.05) {$\mathbf H_K\bw$};
\node[server] (s) at (2.55,0) {server};
\draw[msg] (u1.east) -- node[lab, above] {$X_1$} (u1.east -| s.west);
\draw[msg] (u2.east) -- node[lab, above] {$X_2$} (u2.east -| s.west);
\draw[msg] (uK.east) -- node[lab, above] {$X_K$} (uK.east -| s.west);
\draw[msg] (s.east) -- ++(0.75,0) node[right, font=\small] {$\mathbf D\bw$};
\node[font=\small] at (1.6,-1.85) {(a)};
\end{scope}
\begin{scope}[xshift=6.0cm]
\node[user] (a) at (0,0.85) {$\ba$};
\node[user] (b) at (0,0) {$\bb$};
\node[user] (c) at (0,-0.85) {$\ba+\bb$};
\node[server, minimum height=2.3cm] (t) at (2.4,0) {server};
\draw[msg] (a.east) -- node[lab, above] {$X_1$} (t.west |- 0,0.85);
\draw[msg] (b.east) -- node[lab, above] {$X_2$} (t.west |- 0,0);
\draw[msg] (c.east) -- node[lab, above] {$X_3$} (t.west |- 0,-0.85);
\draw[msg] (t.east) -- ++(0.75,0) node[right, font=\small] {$\ba+\lambda\bb$};
\node[font=\small] at (1.6,-1.85) {(b)};
\end{scope}
\begin{scope}[xshift=12.6cm]
\draw[thick] (-1.1,0) -- (1.1,0) node[right, font=\small] {$(1,0)$};
\draw[thick] (0,-1.1) -- (0,1.1) node[above, font=\small] {$(0,1)$};
\draw[thick] (-0.85,-0.85) -- (0.85,0.85) node[right, font=\small] {$(1,1)$};
\draw[thick, dash pattern=on 4pt off 3pt] (0.55,1.1) -- (-0.55,-1.1) node[below, font=\small, inner sep=1pt] {$(1,\lambda)$};
\node[font=\small] at (0,-1.85) {(c)};
\end{scope}
\end{tikzpicture}
\caption{(a) The problem. (b) The four-line instance. (c) Its observation and demand rows: four lines in a plane, whose cross-ratio is $\lambda$.}\label{fig:model}
\end{figure}

\subsection{Subspace configurations and their invariants}\label{sec:background}

The capacity depends on the coefficient matrices only through a configuration of subspaces. Write $\mathcal H_k=\row(\mathbf H_k)$ and $\mathcal D=\row(\mathbf D)$, subspaces of $\Fq^\mu$, and call the tuple $(\Fq^\mu;\mathcal H_1,\ldots,\mathcal H_K;\mathcal D)$ the \emph{configuration} of the problem. It is a subspace representation of the star quiver with $K+1$ arms~\cite{Gabriel}, and two configurations $(\Fq^\mu;\mathcal H_1,\ldots,\mathcal H_K;\mathcal D)$ and $(\Fq^\mu;\mathcal H'_1,\ldots,\mathcal H'_K;\mathcal D')$ are \emph{equivalent} (isomorphic, in representation theory) if an invertible linear map of $\Fq^\mu$ carries $\mathcal H_k$ to $\mathcal H'_k$ for every $k$ and $\mathcal D$ to $\mathcal D'$. The capacity depends only on the \emph{equivalence class} of the configuration (Section~\ref{sec:invariance}): an invertible change of basis of the data and of each user's observation changes neither what any user observes nor what the server wants. The equivalence class of the configuration is therefore the most fundamental description of the problem: it contains everything the capacity can depend on, and every other invariant is a function of it. Two coarser invariants are commonly used, the first coarser than the second:
\begin{enumerate}
\item the \emph{rank function}, which assigns to each subset of $\{\mathcal H_1,\ldots,\mathcal H_K,\mathcal D\}$ the dimension of the sum of its members, that is, the rank of the stacked coefficient matrices (a point of the linear rank region);
\item the \emph{lattice data}, the dimensions of all subspaces of $\Fq^\mu$ obtained from $\mathcal H_1,\ldots,\mathcal H_K,\mathcal D$ by sums and intersections in any order.
\end{enumerate}
The rank function is a function of the lattice data, which are a function of the equivalence class. The dimension of an intersection of two subspaces is a function of the rank function, for instance $\dim(\mathcal D\cap\mathcal H_1)=\dim\mathcal D+\dim\mathcal H_1-\dim(\mathcal D+\mathcal H_1)$; the lattice data are finer because of intersections of three or more subspaces, such as $\mathcal D\cap\mathcal H_1\cap\mathcal H_2$, and of sums with intersections. In general the lattice data do not determine the equivalence class, and what they miss, such as the cross-ratio of four lines in a plane,\footnote{As in geometry, a line and a plane are subspaces of dimension $1$ and $2$.} is invisible to every rank and to every dimension of sums and intersections; this paper shows that it can determine the capacity. The rank function is exactly what entropy sees: with uniform data, the joint entropy of the users' observations and the demand indexed by a subset is, in $q$-ary units, $L$ times the dimension of the corresponding sum, so the \emph{source entropy vector}, which collects these joint entropies, is $L$ times the rank function. The cut-set bound is a function of the rank function, since each of its terms is the rank of a stacked matrix, and so is every converse derived from the source entropy vector by information inequalities. The capacity formulas known for linear computation problems, such as those of~\cite{Sun_Jafar_CB,Yao_Jafar_3LCBC} and the two-user formula below, are functions of dimensions of sums and intersections, that is, of the lattice data.

The classification of configurations in representation theory says when these invariants determine the equivalence class.\footnote{We state only the facts that frame our results; definitions, proofs and the full classifications are in standard textbooks~\cite{Assem_Simson_Skowronski,Schiffler,Derksen_Weyman} and in the original papers cited here. No result of this paper depends on them.} Three subspaces, as in a problem with two users, are of finite type ($D_4$)~\cite{Gabriel}: the lattice they generate under sums and intersections has at most $28$ elements~\cite{Dedekind}, and the dimensions of these elements determine the equivalence class~\cite{Gelfand_Ponomarev}. Four subspaces, as with three users, are of tame type ($\widetilde D_4$), the four subspace problem of Gelfand and Ponomarev~\cite{Gelfand_Ponomarev}: there are one-parameter families, and the smallest is four distinct lines in a plane, whose lattice data are the same for all such configurations and whose equivalence class is their cross-ratio (Section~\ref{sec:fourline_model}). Five or more subspaces are of wild type~\cite{Nazarova,Brenner}.

In capacity problems the invariants that appear are almost always dimensions of sums and intersections; the field characteristic is known to matter for linear network codes~\cite{Dougherty_Freiling_Zeger}, but that is a dependence between fields. Whether invariants of the configuration finer than its lattice data, within one field, can determine a capacity has rarely been asked. This is the question we study, and we show that they can, and that they matter a great deal: in the smallest case where they exist, they decide whether the cut-set bound is tight or every code must reveal all the data.

\subsection{From two users to the four-line instance}\label{sec:fourline}

We go through the smallest configurations in order of size, to locate the first one where capacity could depend on more than the lattice data.

\emph{One and two users.} With one user the capacity is $\dim \mathcal D$: the user sends the demand. With two users the configuration $(\Fq^\mu;\mathcal H_1,\mathcal H_2;\mathcal D)$ has three subspaces, of finite type, so its equivalence class, and hence the capacity, is a function of the lattice data. The capacity is in fact given by a formula in them, for arbitrary codes (Theorem~\ref{thm:twouser}): user $1$ must send at least $\kappa_1$ symbols per instance, the dimension of the part of the demand that user $2$ does not observe, and user $2$ at least $\kappa_2$, and together they must send another $\kappa_{12}=\dim(\mathcal D\cap \mathcal H_1\cap \mathcal H_2)$ symbols for the component of the demand observed by both, so the capacity is $\kappa_1+\kappa_2+\kappa_{12}$. Here $\kappa_1$ and $\kappa_2$ are functions of the rank function, through intersections of two subspaces, but $\kappa_{12}$, the dimension of an intersection of three, is not. The capacity can strictly exceed the cut-set bound, and the rank function, hence the source entropy vector, does not suffice. If the data are $(w_1,w_2,w_3)$, user $1$ observes $(w_1,w_2)$ and user $2$ observes $(w_1,w_3)$, the demands $(w_2,w_3)$ and $(w_1,w_2+w_3)$ give the same rank function (each space has dimension $2$, and every sum of two or three of them has dimension $3$), but $\kappa_{12}=0$ and $1$, so the capacities are $2$ and $3$, while the cut-set bound, a function of the rank function, is $2$ for both. Two users thus need the lattice data, through the intersection of all three subspaces, and no more. These two configurations are also used in~\cite{Sun_groupcast} to show that extra-entropic structure matters in capacity characterization for secure groupcast.

\emph{Three users with one-dimensional observations.} Let user $k$ observe one symbol $\boldsymbol v_k^\top\bw$ per instance with $\boldsymbol v_k\neq\boldsymbol 0$, so that $\mathcal H_k$ is a line, and let the demand $\mathbf D\bw$ be arbitrary. Up to equivalence there are four cases (Proposition~\ref{prop:threescalar}).
\begin{enumerate}
\item Two of the lines coincide. Users with the same line are merged into one user, whose rate is the sum of theirs, and the two-user formula applies.
\item The three lines $\mathcal H_1,\mathcal H_2,\mathcal H_3$ are distinct and span a space of dimension $3$. After a change of basis the users observe disjoint coordinates, each user sends the part of the demand that involves its own coordinate, and the cut-set bound is tight.
\item The three lines are distinct and span a plane, so that any two users determine the third, and $\mathcal D$ is the whole plane or one of the three users' lines ($\mathcal D=\{\boldsymbol 0\}$ being trivial). The server needs two symbols, or the symbol of one user. In both cases, users sending their observations achieve the cut-set bound.
\item The three lines are distinct and span a plane, and $\mathcal D$ is a fourth line in the plane, distinct from the three users' lines.
\end{enumerate}
The first three cases are settled by the cut-set bound or by the two-user formula, and their answers are functions of dimensions. The fourth case is four distinct lines in a plane. All such configurations have the same rank function, that of the uniform matroid $U_{2,4}$, hence the same source entropy vector, and the same lattice data, but they are not all equivalent: their equivalence class is the cross-ratio of the four lines. So this is the smallest instance in which the capacity could depend on more than the full lattice data. After a change of basis and scaling (Section~\ref{sec:fourline_model}), the users observe $\ba$, $\bb$ and $\ba+\bb$ and the server wants $\ba+\lambda\bb$, where $\lambda\in\Fq\setminus\{0,1\}$ is the cross-ratio; per instance the four symbols range over
\begin{equation}\label{eq:code}
\mathcal Q_\lambda=\{(a,\,b,\,a+b,\,a+\lambda b):\ a,b\in\Fq\},
\end{equation}
the four symbols being the three users' observations and the demand (Figure~\ref{fig:model}(b),(c)). This \emph{four-line instance} is the subject of the paper.

\subsection{Main result and techniques}\label{sec:intro_results}

The cut-set bound for~\eqref{eq:code} says that any two users send at least one symbol per instance in total, so the sum-rate is at least $3/2$, and two users sending their observations, from which the server decodes all the data, gives sum-rate $2$. Write $q=p^n$, and call $\lambda$ \emph{harmonic} if $\lambda=-1$, $\lambda=2$ or $2\lambda=1$. Codes linear over $\Fq$ can do no better than decoding all the data: by an eigenspace argument, they need sum-rate $2$ for every $\lambda$ (Theorem~\ref{thm:linear}). Codes with finer structure can do better. Additive codes, linear only over the prime subfield, reach the cut-set bound when $\lambda\notin\Fp$, while for harmonic $\lambda$ no code at all does better than decoding all the data, and for the remaining $\lambda$ the capacity lies in between. The main result (Theorem~\ref{thm:main}) is this trichotomy in $\lambda$, shown in Figure~\ref{fig:axis}.

\begin{figure}[H]
\centering
\begin{tikzpicture}[font=\small,>={Stealth[length=2mm]},x=1.12cm,y=1.15cm]
\node[anchor=south west] at (0,3.45) {$\lambda\in\Fq\setminus\{0,1\}$};
\draw[rounded corners=5pt] (0,0) rectangle (5.2,3.4);
\node[anchor=west] at (2.85,0.42) {$\lambda\notin\Fp$};
\draw[rounded corners=5pt,fill=gray!10] (0.25,0.85) rectangle (5.0,3.2);
\node[anchor=west] at (0.4,1.27) {$\lambda\in\Fp$, not harmonic};
\draw[rounded corners=5pt,fill=gray!25] (2.4,1.8) rectangle (4.8,3.0);
\node at (3.3,2.4) {harmonic};
\def\ax{6.8}
\draw[->] (\ax,0.0) -- (\ax,3.6) node[above]{sum-rate};
\draw[line width=2.2pt,gray] (\ax,1.1) -- (\ax,3.0);
\draw (\ax-0.08,0.42)--(\ax+0.08,0.42) node[right]{$3/2$, cut-set bound};
\draw (\ax-0.08,1.1)--(\ax+0.08,1.1) node[right]{$3p/(2p-1)$, sumset bound};
\draw (\ax-0.08,3.0)--(\ax+0.08,3.0) node[right]{$2$, decode all};
\node[anchor=west] at (\ax+0.12,2.5) {also $\Fq$-linear codes, every $\lambda$};
\node[anchor=west,gray!70!black] at (\ax+0.12,1.85) {open};
\fill (\ax,0.42) circle (2.2pt); \fill (\ax,3.0) circle (2.2pt);
\draw[thick] (\ax-0.15,1.1) -- (\ax,1.1); \draw[thick] (\ax-0.15,3.0) -- (\ax,3.0);
\fill (4.4,0.42) circle (1.3pt); \draw[->,thick] (4.4,0.42) -- (\ax-0.15,0.42);
\fill (4.3,1.27) circle (1.3pt); \draw[->,densely dashed] (4.3,1.27) -- (\ax-0.18,1.85);
\fill (4.5,2.4) circle (1.3pt); \draw[->,thick] (4.5,2.4) -- (\ax-0.15,2.95);
\end{tikzpicture}
\caption{The cross-ratios divided into three cases. Solid arrows point to the capacity, which is determined in two cases; the dashed arrow points to the interval known to contain the capacity in the third.}\label{fig:axis}
\end{figure}

\begin{enumerate}
\item \emph{$\lambda\notin\Fp$: capacity $3/2$.} The cut-set bound is achieved by codes that are linear over $\Fp$ but not over $\Fq$, built from the Frobenius map $z\mapsto z^p$, with block length $2$ for every $q$ (Theorem~\ref{thm:additive}).
\item \emph{$\lambda$ harmonic: capacity $2$.} No code does better than decoding all the data: by a confusability argument, every zero-error code reveals all the data to the server (Theorem~\ref{thm:harmonic}).
\item \emph{$\lambda\in\Fp$, not harmonic: capacity in $[3p/(2p-1),2]$.} No code approaches the cut-set bound at any block length, by a sumset bound from additive combinatorics, the Pl\"unnecke--Ruzsa inequality~\cite{Ruzsa,Petridis} (Theorem~\ref{thm:plunnecke}). This case is empty for $p\in\{2,3,5\}$, so over fields of characteristic $2$, $3$ and $5$ the capacity is determined for every $\lambda$ (Corollary~\ref{cor:complete}); the first open case is $p=7$, $\lambda\in\{3,5\}$, with capacity in $[12/7,2]$ (Remark~\ref{rem:p7}).\footnote{The nonlinear repair schemes of~\cite{Con_Tamo}, applied to the four-line instance, show that the capacity is below $2$ for almost all such $\lambda$ as $p\to\infty$; they do not decide a particular case, such as $p=7$, $\lambda\in\{3,5\}$.}
\end{enumerate}
Each technique builds on a known tool, linear coding, confusability or sumset inequalities, and what is new is how the arithmetic of the configuration enters it.

\emph{Consequences.} Over $\F_{3^2}$, the four-line instances with $\lambda=-1$ and with $\lambda\notin\F_3$ have the same rank function, hence the same source entropy vector, and the same lattice data, but their capacities are $2$ and $3/2$ (Corollary~\ref{cor:lattice}). So no formula in the lattice data, with sums and intersections taken in any order, determines the capacity; it depends on the instance through its cross-ratio and the prime subfield. To our knowledge, the cross-ratio has not previously appeared as a quantity that determines a capacity. The same pair shows that the source entropy vector does not capture the capacity: the encoding and decoding dependencies are the same entropy equalities for both, and the $\F_{3^2}$ code for $\lambda\notin\F_3$ gives an entropic vector that satisfies all of them with sum-rate $3/2$. So every converse that uses only these entropies, these dependencies and inequalities valid for entropic vectors, Shannon or not, gives at most $3/2$ for $\lambda=-1$, where the capacity is $2$; the confusability argument uses the code itself, not only these entropies (Remark~\ref{rem:entropic}). The four-line instance thus shows three phenomena at once: capacity is not a function even of the full lattice data; coding linear over the prime subfield beats coding linear over $\Fq$; and no information inequality applied to the source entropy vector and the dependencies alone yields the converse.

\emph{Vanishing error.} The qualitative separation persists under vanishing error for uniform data: the capacity is still the cut-set bound $3/2$ exactly when $\lambda\notin\Fp$; for every $\lambda\in\Fp$ it is still at least $3(3p-1)/(6p-4)>3/2$; and codes linear over $\Fq$ or over $\Fp$ achieve the same sum-rates as with zero error (Theorem~\ref{thm:ve}).

\subsection{Connection to repair}

The four symbols of~\eqref{eq:code} range over a scalar linear $[4,2]$ MDS code, and every such code is of this form (Section~\ref{sec:fourline_model}), so the four-line instance is also exact repair of one node of a $[4,2]$ storage code from the other three, with $L$ codewords repaired jointly; the cut-set bound is that of~\cite{Dimakis_etal}. The repair literature studies mainly schemes linear over a subfield~\cite{Shanmugam_Papailiopoulos_Dimakis_Caire,Guruswami_Wootters,Tamo_Ye_Barg}, which for $[4,2]$ codes achieve the cut-set bound when $n$ is even and $\lambda\notin\F_{p^{n/2}}$~\cite{Shanmugam_Papailiopoulos_Dimakis_Caire}, and nonlinear schemes over prime fields~\cite{Con_Tamo}, which approach it for almost all cross-ratios as $p\to\infty$. Recent works study repair of Reed--Solomon codes at the cut-set bound~\cite{Fang_Qiu_Xia_Fu,Li_cutset}. Appendix~\ref{app:repair} translates our results and compares them with the repair literature.

\section{Problem Statement}\label{sec:model}

This section states the general problem, the invariance that reduces it to subspace configurations, and the four-line instance.

\subsection{Capacity, linear capacity and additive capacity}

There are $K$ users and one server. Vectors are columns. Fix full row rank matrices $\mathbf H_k\in\Fq^{\mu_k\times\mu}$, $k\in[K]$, and $\mathbf D\in\Fq^{\mu_0\times\mu}$ with $\row(\mathbf D)\subseteq\row(\mathbf H_1)+\cdots+\row(\mathbf H_K)$. A code of block length $L$ operates on $L$ instances, whose data form the vector $\bw^L=(\bw_1^\top,\ldots,\bw_L^\top)^\top\in\Fq^{\mu L}$, uniform over $\Fq^{\mu L}$, where $\bw_\ell\in\Fq^\mu$ is the data of instance $\ell$; a value of $\bw^L$ is an \emph{input}. User $k$'s observation over the $L$ instances is $\mathbf H_k^{[L]}\bw^L=((\mathbf H_k\bw_1)^\top,\ldots,(\mathbf H_k\bw_L)^\top)^\top$, where $\mathbf H_k^{[L]}=\mathrm{diag}(\mathbf H_k,\ldots,\mathbf H_k)$ is block diagonal with $L$ blocks. From this observation, user $k$ sends a message $X_k=\phi_k(\mathbf H_k^{[L]}\bw^L)$ to the server, taking values in its \emph{message alphabet} $\cX_k$. The server outputs $\psi(X_1,\ldots,X_K)$ and the code is \emph{zero-error} if $\psi(X_1,\ldots,X_K)=\mathbf D^{[L]}\bw^L$ for every $\bw^L$, with $\mathbf D^{[L]}$ defined in the same way; Appendix~\ref{app:ve} treats vanishing error. User $k$ sends $N_k=\log_q|\cX_k|$ $q$-ary symbols, and its rate is $R_k=N_k/L$. A rate tuple $(R_1,\ldots,R_K)$ is \emph{achievable} if for some $L$ there is a zero-error code whose rates are at most $R_1,\ldots,R_K$, and the \emph{capacity region} $\mathcal C$ is the closure of the set of achievable rate tuples. The (sum) capacity is the minimum of $R_1+\cdots+R_K$ over $\mathcal C$. Throughout, $q=p^n$ with $p$ prime, and $\Fp\subseteq\Fq$ is the prime subfield.

\emph{Linear and additive codes.} A code is \emph{$\Fq$-linear} if its encoders and decoder are matrices over $\Fq$: user $k$ sends $X_k=\mathbf F_k\mathbf H_k^{[L]}\bw^L\in\Fq^{N_k}$ with $\mathbf F_k\in\Fq^{N_k\times\mu_kL}$, so that $N_k$ is an integer, and the server computes $\mathbf D^{[L]}\bw^L=\mathbf U_1X_1+\cdots+\mathbf U_KX_K$ with $\mathbf U_k\in\Fq^{\mu_0L\times N_k}$.\footnote{\label{fn:linear-decoder}Linear encoders with an arbitrary decoder are no more general. Stack the messages as $(X_1,\ldots,X_K)=\mathbf G\bw^L$ with $\mathbf G$ the matrix whose blocks are $\mathbf F_k\mathbf H_k^{[L]}$. Zero error says that $\mathbf G\bw^L=\mathbf G\bw'^L$ implies $\mathbf D^{[L]}\bw^L=\mathbf D^{[L]}\bw'^L$, that is, every $\bw^L$ with $\mathbf G\bw^L=\boldsymbol 0$ has $\mathbf D^{[L]}\bw^L=\boldsymbol 0$; so the rows of $\mathbf D^{[L]}$ lie in the row space of $\mathbf G$, and $\mathbf D^{[L]}=\mathbf U\mathbf G$ for a matrix $\mathbf U$. The same holds over $\Fp$ for additive codes.} A code is \emph{additive}, or $\Fp$-linear, if after identifying $\Fq$ with the vector space $\Fp^n$ through a fixed basis, its encoders and decoder are matrices over $\Fp$: user $k$ sends $X_k\in\Fp^{nN_k}$ obtained by applying a matrix $\mathbf F_k\in\Fp^{nN_k\times n\mu_kL}$ to its observation expressed over $\Fp$, so that $nN_k$ is an integer. Multiplication by a fixed $\gamma\in\Fq$ acts on the $n$ coordinates of a symbol as an $n\times n$ matrix over $\Fp$, so an $\Fq$-linear code is additive, with matrices made of such $n\times n$ blocks. An additive code need not have this block form, so for $n\geq 2$ it is more general, as the smallest field shows below.

\begin{example}[Linear and additive maps over $\F_{2^2}$]\label{ex:linadd}
Let $q=2^2$, $\F_{2^2}=\{0,1,\omega,\omega^2\}$ with $\omega^2=\omega+1$, identified with $\F_2^2$ through the basis $1,\omega$, so that $a=a_0+\omega a_1$ corresponds to $(a_0,a_1)^\top\in\F_2^2$. An $\F_{2^2}$-linear map from one symbol of $\F_{2^2}$ to one symbol of $\F_{2^2}$ is multiplication by some $\gamma\in\F_{2^2}$; since $\omega a=a_1+\omega(a_0+a_1)$ and $\omega^2a=(a_0+a_1)+\omega a_0$, the four choices $\gamma=0,1,\omega,\omega^2$ act on $(a_0,a_1)^\top$ as the matrices
\begin{equation*}
\begin{bmatrix}0&0\\0&0\end{bmatrix},\quad\begin{bmatrix}1&0\\0&1\end{bmatrix},\quad\begin{bmatrix}0&1\\1&1\end{bmatrix},\quad\begin{bmatrix}1&1\\1&0\end{bmatrix}
\end{equation*}
over $\F_2$. An additive map is an $\F_2$-linear map on $\F_2^2$: to $\F_2^2$ it is any of the $16$ matrices in $\F_2^{2\times 2}$, and to $\F_2$ any row vector in $\F_2^{1\times 2}$. Two such additive maps that are not $\F_{2^2}$-linear stand out. The squaring map $a\mapsto a^2$ gives $(a_0+\omega a_1)^2=(a_0+a_1)+\omega a_1$, since the cross term $2\omega a_0a_1$ vanishes in characteristic $2$, $a_0^2=a_0$ and $a_1^2=a_1$ for bits, and $\omega^2=1+\omega$; its matrix is $\begin{bmatrix}1&1\\0&1\end{bmatrix}$, which is not among the four above. The map $a\mapsto a_0$, with matrix $(1\ \,0)$, sends half a symbol of $\F_{2^2}$, which no $\F_{2^2}$-linear map can do, since its messages consist of whole symbols of $\F_{2^2}$.
\end{example}

\emph{Matrices and linearized polynomials.} Example~\ref{ex:linadd} writes an additive map in two ways: as a matrix over $\Fp$ acting on coordinates, and as a polynomial over $\Fq$, such as $a\mapsto a^2$. Both representations are general. After identifying $\Fq$ with $\Fp^n$, an additive map from $\Fq$ to $\Fq$ is represented by a matrix in $\Fp^{n\times n}$; it also has a unique representation $z\mapsto\gamma_0z+\gamma_1z^p+\cdots+\gamma_{n-1}z^{p^{n-1}}$ with $\gamma_i\in\Fq$, called a \emph{linearized polynomial}~\cite[Chapter~3]{Lidl_Niederreiter}. The $\Fq$-linear maps are those with only the term $\gamma_0z$. The translation between the two representations depends on the basis and is not transparent: squaring is the matrix $\left[\begin{smallmatrix}1&1\\0&1\end{smallmatrix}\right]$ in Example~\ref{ex:linadd}, from which it is not apparent that squaring fixes exactly the elements of $\F_2$, and the matrix of a general linearized polynomial is harder still to read. We therefore use whichever representation is cleaner: matrix representations for the definitions, and polynomial representations for the codes and proofs, where a power $z^p$ makes the structure visible. The two representations describe the same additive maps, and a zero-error decoder for additive encoders can be taken to be $\Fp$-linear by Footnote~\ref{fn:linear-decoder}, whether or not its matrix is written out.

The \emph{linear capacity region} $\mathcal C_{\mathrm{lin}}$ and the \emph{additive capacity region} $\mathcal C_{\mathrm{add}}$ are obtained by restricting the definition of $\mathcal C$ to $\Fq$-linear codes and additive codes, respectively. Their minimum sum-rates are the \emph{linear capacity} and the \emph{additive capacity}. Since every $\Fq$-linear code is additive, $\mathcal C_{\mathrm{lin}}\subseteq\mathcal C_{\mathrm{add}}\subseteq\mathcal C$.

The \emph{cut-set bound} is the converse obtained by giving the server the observations of the users outside a set $\mathcal S\subseteq[K]$: for every $\mathcal S$,
\begin{equation}\label{eq:cutset}
\sum_{k\in\mathcal S}R_k\ \geq\ \frac{1}{L}H(\mathbf D^{[L]}\bw^L\,|\,\mathbf H_k^{[L]}\bw^L,\ k\notin\mathcal S)=\rank\begin{bmatrix}\mathbf D\\ \mathbf H_{\mathcal S^c}\end{bmatrix}-\rank\mathbf H_{\mathcal S^c},
\end{equation}
where $\mathbf H_{\mathcal S^c}$ stacks $\mathbf H_k$, $k\notin\mathcal S$, and entropies are in $q$-ary units throughout. The users in $\mathcal S$ must supply all information about the demand that is not already contained in the observations of the users outside $\mathcal S$.

\subsection{Invariance, ranks and lattice data}\label{sec:invariance}

An invertible matrix $\mathbf P\in\Fq^{\mu\times\mu}$ changes the coordinates of the data: writing $\bw=\mathbf P\widetilde{\bw}$ replaces $\mathbf H_k$ and $\mathbf D$ by $\mathbf H_k\mathbf P$ and $\mathbf D\mathbf P$, and the transformed data $\widetilde{\bw}$ remain uniform. Separately, invertible matrices $\mathbf Q_k\in\Fq^{\mu_k\times\mu_k}$ and $\mathbf Q\in\Fq^{\mu_0\times\mu_0}$ change the coordinates of user $k$'s observation and of the demand, respectively. Each user can invert the change to its observation, and recovering $\mathbf Q\mathbf D\bw$ is equivalent to recovering $\mathbf D\bw$. Thus replacing $\mathbf H_k$ by $\mathbf Q_k\mathbf H_k\mathbf P$ and $\mathbf D$ by $\mathbf Q\mathbf D\mathbf P$ preserves $\mathcal C$. Left multiplication by $\mathbf Q_k$ and $\mathbf Q$ leaves the row spaces unchanged, while right multiplication by $\mathbf P$ applies a common invertible map to all row spaces. Hence $\mathcal C$ depends only on the equivalence class of the configuration $(\Fq^\mu;\mathcal H_1,\ldots,\mathcal H_K;\mathcal D)$, where $\mathcal H_k=\row(\mathbf H_k)$ and $\mathcal D=\row(\mathbf D)$.

As in Section~\ref{sec:background}, the rank function records the dimension of the sum of each subset of $\mathcal H_1,\ldots,\mathcal H_K,\mathcal D$. For uniform data, the joint entropy of any subset of the observations $\mathbf H_1^{[L]}\bw^L,\ldots,\mathbf H_K^{[L]}\bw^L$ and the demand $\mathbf D^{[L]}\bw^L$ equals $L$ times the dimension of the corresponding sum. Thus the source entropy vector is $L$ times the rank function, and the cut-set bound~\eqref{eq:cutset} depends only on the rank function. The lattice data record the dimensions of all subspaces generated from $\mathcal H_1,\ldots,\mathcal H_K,\mathcal D$ by sums and intersections in any order, and therefore determine the rank function. For $K\leq 2$, the classification recalled in Section~\ref{sec:background} shows that the lattice data also determine the equivalence class, and hence the capacity region; Theorem~\ref{thm:twouser} gives the region explicitly. For $K=3$, the four-line instance is the first configuration whose equivalence class is not determined by the lattice data.

\subsection{The four-line instance}\label{sec:fourline_model}

Set $\mu=2$, $K=3$, $\mathbf H_1=(1,0)$, $\mathbf H_2=(0,1)$, $\mathbf H_3=(1,1)$ and $\mathbf D=(1,\lambda)$ with $\lambda\in\Fq\setminus\{0,1\}$; $q\geq 3$. Writing $\bw_\ell=(a_\ell,b_\ell)^\top$, $\ba=(a_1,\ldots,a_L)^\top$ and $\bb=(b_1,\ldots,b_L)^\top$, user 1 observes $\ba$, user 2 observes $\bb$, user 3 observes
\begin{equation}\label{eq:c}
\bs=\ba+\bb,
\end{equation}
and the server wants
\begin{equation}\label{eq:w}
\by=\ba+\lambda\bb.
\end{equation}

Every configuration of four distinct lines in a plane is equivalent to a four-line instance. After a change of basis of the data the first two users observe $w_1$ and $w_2$, the third observes $\zeta_1w_1+\zeta_2w_2$ and the server wants $\eta_1w_1+\eta_2w_2$, and the four lines are distinct exactly when $\zeta_1\zeta_2\eta_1\eta_2\neq 0$ and $\zeta_1\eta_2-\zeta_2\eta_1\neq 0$. With $a=\zeta_1w_1$ and $b=\zeta_2w_2$, the third user observes $a+b$, and the demand multiplied by $\zeta_1/\eta_1$ is $a+\lambda b$ with
\begin{equation*}
\lambda=\frac{\zeta_1\eta_2}{\zeta_2\eta_1}\notin\{0,1\},
\end{equation*}
the \emph{cross-ratio} of the four lines. The cross-ratio determines the equivalence class, and conversely: an invertible map of $\Fq^2$ that carries each of the lines spanned by $(1,0)$, $(0,1)$ and $(1,1)$ to itself is diagonal by the first two and a scalar by the third, so it fixes every line, and the lines spanned by $(1,\lambda)$ and $(1,\lambda')$ give equivalent configurations only if $\lambda=\lambda'$. This is the classical fact that the cross-ratio is the only invariant of four distinct points of the projective line~\cite{Coxeter}. The lattice data do not see it: any two of the four lines intersect in $\{\boldsymbol 0\}$ and sum to the plane, so sums and intersections in any order give only $\{\boldsymbol 0\}$, the four lines and $\Fq^2$, with dimensions $0$, $1$ and $2$, for every $\lambda$. Equivalently, the four symbols range over a scalar linear $[4,2]$ MDS code, and every such code is $\mathcal Q_\lambda$ up to scaling of its coordinates and a change of basis.

Any two of $\ba,\bb,\bs,\by$ determine the other two. The four vectors are uniform on $\Fq^L$ and pairwise independent. The three messages form the \emph{message triple} $(X_1,X_2,X_3)$. For this instance, the cut-set bound~\eqref{eq:cutset} is
\begin{equation}\label{eq:cutset4}
R_1+R_2\geq 1,\quad R_1+R_3\geq 1,\quad R_2+R_3\geq 1,
\end{equation}
with the singleton sets giving $R_k\geq 0$. We call
\begin{equation*}
\cR_{\mathrm{cut}}=\{(R_1,R_2,R_3)\in\mathbb R_+^3:\ R_i+R_j\geq 1\ \text{for all}\ i\neq j\}
\end{equation*}
the \emph{cut-set region}; its vertices are $(1/2,1/2,1/2)$ and the permutations of $(1,1,0)$. We call
\begin{equation*}
\cR_{\mathrm{all}}=\{(R_1,R_2,R_3)\in\mathbb R_+^3:\ R_i+R_j\geq 1\ \text{for all}\ i\neq j,\ R_1+R_2+R_3\geq 2\}
\end{equation*}
the \emph{decode-all region}; its vertices are the permutations of $(1,1,0)$, so it is the upward closure $\mathrm{conv}\{(1,1,0),(1,0,1),(0,1,1)\}+\mathbb R_+^3$ of the three schemes in which two users send their observations and the server decodes all the data.

Relabeling the users, that is, permuting the roles of $\ba,\bb,\bs$, followed by scaling replaces $\lambda$ by one of the six values $\lambda,1-\lambda,1/\lambda,1/(1-\lambda),(\lambda-1)/\lambda,\lambda/(\lambda-1)$, which we call the \emph{relabelings} of $\lambda$, and permutes the rate coordinates accordingly. Hence cross-ratios that are relabelings of one another have regions $\mathcal C$, $\mathcal C_{\mathrm{lin}}$, $\mathcal C_{\mathrm{add}}$ that agree up to the corresponding permutation of coordinates. All fully characterized capacity regions for the four-line instance are symmetric in the three users, so for them the region depends on $\lambda$ only up to relabeling.\footnote{For instance, $\lambda=2$ becomes $\lambda=-1$ by letting user 3 play the role of user 1: with $\widetilde{\bb}=-\bb$, the users observe $\bs$, $\widetilde{\bb}$ and $\ba=\bs+\widetilde{\bb}$, and the server wants $\ba+2\bb=\bs-\widetilde{\bb}$.}

We call $\lambda$ harmonic, as in projective geometry, if $\lambda=-1$, $\lambda=2$ or $2\lambda=1$; for odd $q$ these three values are relabelings of one another and lie in $\Fp$, and for even $q$ no $\lambda\notin\{0,1\}$ is harmonic.

\section{Main results: the four-line instance}\label{sec:results}

We begin with the two-user capacity region and the classification of three users with one-dimensional observations, which justify the reduction in Section~\ref{sec:fourline}. We then state the main theorem for the four-line instance and develop its three cases in order, followed by the linear and additive capacity regions and the consequences for rank, lattice data and entropy arguments.

\subsection{Warm-up: two users, and three users with one-dimensional observations}\label{sec:warmup}

For two users, the lattice data determine the capacity region for arbitrary codes. With $\mathcal H_1$, $\mathcal H_2$ and $\mathcal D\subseteq\mathcal H_1+\mathcal H_2$ as in Section~\ref{sec:model}, define
\begin{equation*}
\kappa_1=\dim \mathcal D-\dim(\mathcal D\cap \mathcal H_2),\qquad \kappa_2=\dim \mathcal D-\dim(\mathcal D\cap \mathcal H_1),\qquad \kappa_{12}=\dim(\mathcal D\cap \mathcal H_1\cap \mathcal H_2).
\end{equation*}

\begin{theorem}[Two users]\label{thm:twouser}
For $K=2$, $\mathcal C=\mathcal C_{\mathrm{lin}}=\{(R_1,R_2):\ R_1\geq\kappa_1,\ R_2\geq\kappa_2,\ R_1+R_2\geq\kappa_1+\kappa_2+\kappa_{12}\}$.
\end{theorem}

\begin{remark}[Gap to the cut-set bound]\label{rem:gap}
The individual bounds coincide with the single-user cut-set bounds. Together with $R_1+R_2\geq\dim\mathcal D$, they give the cut-set lower bound $\max\{\dim\mathcal D,\kappa_1+\kappa_2\}$ on the sum-rate. Let $\kappa_0=\dim\mathcal D-\dim\big((\mathcal D\cap\mathcal H_1)+(\mathcal D\cap\mathcal H_2)\big)$. The capacity is $\kappa_1+\kappa_2+\kappa_{12}=\dim\mathcal D+\kappa_0$, so the gap above the cut-set bound is $\min\{\kappa_0,\kappa_{12}\}$. The gap is therefore positive exactly when $\kappa_0>0$ and $\kappa_{12}>0$, that is, when $\mathcal D\neq(\mathcal D\cap\mathcal H_1)+(\mathcal D\cap\mathcal H_2)$ and $\mathcal D\cap\mathcal H_1\cap\mathcal H_2\neq\{\boldsymbol 0\}$. For example, suppose the users observe $(w_1,w_0)$ and $(w_2,w_0)$ and the server wants $(w_1+w_2,w_0)$. Then $\kappa_1=\kappa_2=\kappa_{12}=1$, and the capacity region is $R_1\geq1$, $R_2\geq1$, $R_1+R_2\geq3$, whereas the cut-set bound allows sum-rate $2$.
\end{remark} 

The next proposition gives the classification used in Section~\ref{sec:fourline} for three users with one-dimensional observations. The proofs of Theorem~\ref{thm:twouser} and Proposition~\ref{prop:threescalar} are in Appendix~\ref{app:twouser}.

\begin{proposition}[Three users with one-dimensional observations]\label{prop:threescalar}
Let $K=3$ and $\mathbf H_k=\boldsymbol v_k^\top$ with $\boldsymbol v_k\neq\boldsymbol 0$. If two of the $\boldsymbol v_k$ are proportional, the capacity region follows from Theorem~\ref{thm:twouser} by merging users with the same $\mathcal H_k$ into one user whose rate is the sum of theirs. If the $\boldsymbol v_k$ are pairwise independent, the capacity region is the cut-set region, except when they span a plane and $\mathcal D$ is a fourth line in it, which is the four-line instance.
\end{proposition}

\subsection{The main theorem}

For the four-line instance~\eqref{eq:code}, the following theorem determines the capacity region except when $\lambda$ is non-harmonic and lies in the prime subfield. The regions $\cR_{\mathrm{cut}}$ and $\cR_{\mathrm{all}}$ are defined in Section~\ref{sec:fourline_model}.

\begin{theorem}[Four-line capacity]\label{thm:main}
Let $q=p^n$ and $\lambda\in\Fq\setminus\{0,1\}$.
\begin{enumerate}
\item If $\lambda\notin\Fp$, then $\mathcal C=\mathcal C_{\mathrm{add}}=\cR_{\mathrm{cut}}$, and the capacity is $3/2$.
\item If $q$ is odd and $\lambda$ is harmonic, then $\mathcal C=\cR_{\mathrm{all}}$, the capacity is $2$, and in every zero-error code the message triple determines $(\ba,\bb)$.
\item If $\lambda\in\Fp$ is not harmonic, then $\mathcal C_{\mathrm{add}}=\cR_{\mathrm{all}}\subseteq\mathcal C\subsetneq\cR_{\mathrm{cut}}$, and the capacity lies in $[3p/(2p-1),2]$.
\end{enumerate}
In every case $\mathcal C_{\mathrm{lin}}=\cR_{\mathrm{all}}$.
\end{theorem}

The first two cases follow from Theorems~\ref{thm:additive} and~\ref{thm:harmonic}. In the third case, Corollary~\ref{cor:additive} gives $\mathcal C_{\mathrm{add}}=\cR_{\mathrm{all}}$, and Theorem~\ref{thm:plunnecke} gives $\mathcal C\subsetneq\cR_{\mathrm{cut}}$ and the lower bound $3p/(2p-1)$. Two users sending their observations achieve the upper bound $2$. Theorem~\ref{thm:linear} gives the linear capacity region. Table~\ref{tab:summary} summarizes the three cases.

The third case is empty for $p\in\{2,3,5\}$: $\F_2\setminus\{0,1\}$ is empty, $\F_3\setminus\{0,1\}=\{2\}$ with $2=-1$, and $\F_5\setminus\{0,1\}=\{2,3,4\}$ with $3=1/2$ and $4=-1$. Thus every cross-ratio in the prime subfield is harmonic in these characteristics, giving the following complete characterization.

\begin{corollary}[Characteristics two, three and five]\label{cor:complete}
If $p\in\{2,3,5\}$, the capacity region of the four-line instance over $\F_{p^n}$ is $\cR_{\mathrm{cut}}$ for $\lambda\notin\Fp$ and $\cR_{\mathrm{all}}$ for $\lambda\in\Fp\setminus\{0,1\}$.
\end{corollary}

\begin{table}[H]
\centering
\begin{tabular}{lcccc}
\toprule
$\lambda$ & $\mathcal C$ & $\mathcal C_{\mathrm{add}}$ & $\mathcal C_{\mathrm{lin}}$ & capacity\\
\midrule
$\lambda\notin\Fp$ & $\cR_{\mathrm{cut}}$ & $\cR_{\mathrm{cut}}$ & $\cR_{\mathrm{all}}$ & $3/2$\\
$\lambda$ harmonic, $q$ odd & $\cR_{\mathrm{all}}$ & $\cR_{\mathrm{all}}$ & $\cR_{\mathrm{all}}$ & $2$\\
$\lambda\in\Fp$, not harmonic & open & $\cR_{\mathrm{all}}$ & $\cR_{\mathrm{all}}$ & in $[3p/(2p-1),2]$\\
\bottomrule
\end{tabular}
\caption{The four-line instance.}\label{tab:summary}
\end{table}

\subsection{Cross-ratios outside the prime subfield}\label{sec:prime-free}

Outside the prime subfield, additive codes achieve the entire cut-set region.

\begin{theorem}[Cross-ratios outside $\Fp$]\label{thm:additive}
Let $\lambda\in\Fq\setminus\Fp$. Then $\mathcal C=\mathcal C_{\mathrm{add}}=\cR_{\mathrm{cut}}$. In particular, the rate tuple $(1/2,1/2,1/2)$, at which every user sends half its observation, is achieved by an additive code with $L=2$.
\end{theorem}

The proof is in Section~\ref{sec:proof_additive}. The code hides changes of the data that cancel in the demand, and the position of $\lambda$ relative to the prime subfield decides whether these hidden changes can be kept apart from all others. A change $(\Delta\ba,\Delta\bb)=(-\lambda\bz,\bz)$ gives $\Delta\by=\Delta\ba+\lambda\Delta\bb=\boldsymbol 0$, and it changes the three observations by $-\lambda\bz$, $\bz$ and $(1-\lambda)\bz$. This suggests taking an $\Fp$-subspace $U\subseteq\Fq^2$ of half the ambient dimension and additive encoders with null spaces $\lambda U$, $U$ and $(1-\lambda)U$ for users $1$, $2$ and $3$: each user sends half its observation, giving the rate tuple $(1/2,1/2,1/2)$, and every change $(-\lambda\bz,\bz)$ with $\bz\in U$ leaves all three messages unchanged. Each encoder also hides all other changes in its null space, so zero error needs a further condition. The condition $U\cap\lambda U=\{\boldsymbol 0\}$, together with $\bs=\ba+\bb$, forces every change preserving the three messages to have the form $(-\lambda\bz,\bz)$ for one $\bz\in U$. Hence the demand is unchanged.

The prime subfield decides whether such a $U$ exists. Multiplication by a nonzero element of $\Fp$ maps every $\Fp$-subspace onto itself, so $\lambda U=U$ when $\lambda\in\Fp$, and no nonzero $U$ satisfies the condition. Multiplication by $\lambda\notin\Fp$ need not preserve $U$, and can move it to a subspace that meets $U$ only in $\boldsymbol 0$. A subspace built from the Frobenius map $z\mapsto z^p$ does this, since the fixed points of this map are exactly the elements of $\Fp$; Section~\ref{sec:proof_additive} gives it and the three encoders explicitly and shows that they are additive. They are not $\Fq$-linear, since by Theorem~\ref{thm:linear} every $\Fq$-linear code has sum-rate at least $2$.

\subsection{Harmonic cross-ratios}

For harmonic $\lambda$, no code improves on decoding all the data. The converse is structural: in every zero-error code, whatever its rates, the message triple determines $(\ba,\bb)$.

\begin{theorem}[Harmonic cross-ratios]\label{thm:harmonic}
Let $q$ be odd and $\lambda$ be harmonic. In every zero-error code, $(X_1,X_2,X_3)$ determines $(\ba,\bb)$. Consequently $\mathcal C=\cR_{\mathrm{all}}$.
\end{theorem}

The proof is in Section~\ref{sec:proof_harmonic}. Call two inputs \emph{confusable} if they produce the same message triple. After relabeling and scaling, take $\lambda=-1$. If two distinct inputs $(\ba,\bb)$ and $(\ba',\bb')$ were confusable, zero error would give them the same demand, $\ba-\bb=\ba'-\bb'$. Exchange the observations of user $2$, which gives the inputs $(\ba,\bb')$ and $(\ba',\bb)$. Users $1$ and $2$ send the same messages as before, and the sums $\ba+\bb'$ and $\ba'+\bb$ are equal by $\ba-\bb=\ba'-\bb'$, so user $3$ sends equal messages on the two exchanged inputs, possibly different from its original message. Hence the exchanged inputs are also confusable. Their demands $\ba-\bb'$ and $\ba'-\bb$, however, differ, since $q$ is odd and the inputs are distinct, contradicting zero error. Therefore no two distinct inputs are confusable, and the message triple determines $(\ba,\bb)$.

\begin{remark}[Where the harmonic cross-ratio enters]
The exchange is where $\lambda=-1$ enters. For general $\lambda$, two confusable inputs with the same demand are $(\ba,\bb)$ and $(\ba-\lambda\bz,\bb+\bz)$ with $\bz\neq\boldsymbol 0$, and the exchanged inputs $(\ba,\bb+\bz)$ and $(\ba-\lambda\bz,\bb)$ have sums $\bs+\bz$ and $\bs-\lambda\bz$, which are equal only when $\lambda=-1$. Otherwise user 3 observes different sums, so its messages need not agree and the exchanged inputs need not be confusable; the code of Theorem~\ref{thm:additive} exploits this freedom.
\end{remark}

\subsection{Cross-ratios in the prime subfield}\label{sec:prime}

For cross-ratios in the prime subfield, every code has sum-rate bounded away from $3/2$. To explain it, identify $\Fp$ with $\{0,1,\ldots,p-1\}$, so that elements of $\Fp$ can also be compared and used as integers, and for $z\in\Fp\setminus\{0\}$ let $\rho(z)=1/z\in\{1,\ldots,p-1\}$; then $\rho(z)\geq2$ when $z\neq1$.

\begin{theorem}[Cross-ratios in $\Fp$]\label{thm:plunnecke}
Let $\lambda\in\Fp\setminus\{0,1\}$ and $q=p^n$. Every zero-error code of every block length $L$ satisfies
\begin{equation}\label{eq:plun_sum}
R_1+R_2+R_3\ \geq\ \frac{3p}{2p-1}\ >\ \frac32.
\end{equation}
\end{theorem}

The proof is in Section~\ref{sec:proof_plunnecke}. Theorem~\ref{thm:plunnecke} is a robust form of the obstruction in Section~\ref{sec:prime-free}: the construction there needs an $\Fp$-subspace $U$ with $U\cap\lambda U=\{\boldsymbol 0\}$, which is impossible for $\lambda\in\Fp$ because then $\lambda U=U$. A general code need not use subspaces, and the theorem says that the obstruction survives, in approximate form, for arbitrary sets.

These sets are the bins of the code. The \emph{bin} of a message of user $k$ is the set of observations that $\phi_k$ maps to that message; let $\cA$ and $\cB$ be largest bins of users~1 and~2, which are large when these users send little. On $\cA\times\cB$ the messages of users~1 and~2 are fixed. Two pairs with the same sum $\ba+\bb$ then have the same message triple and, by zero error, the same demand, and the sum and the demand determine the pair. So the $|\cA||\cB|$ sums are distinct. The demands $\cA+\lambda\cB$, in contrast, are determined by the message of user~3, so they take few values when user~3 sends little. Few values of $\cA+\lambda\cB$ mean that adding $\lambda\cB$ barely enlarges $\cA$. When $\lambda\in\Fp$, every $\bb$ is an integer multiple of $\lambda\bb$, namely $\bb=\rho(\lambda)\cdot\lambda\bb$, so $\cB$ is contained in the sum of $\rho(\lambda)$ copies of $\lambda\cB$. The Pl\"unnecke--Ruzsa inequality, a basic tool of additive combinatorics, says that adding repeatedly a set that barely enlarges $\cA$ still barely enlarges a large part of $\cA$. Then adding $\cB$ would barely enlarge that part, contradicting that these sums are distinct. Quantitatively, $|\cA+\lambda\cB|\geq|\cA||\cB|^{1/\rho(\lambda)}$. With the rates this gives $R_1+R_3+R_2/\rho(\lambda)\geq 1+1/\rho(\lambda)$. Relabeling gives one such inequality for each user, and adding them, with the cut-set bounds, gives~\eqref{eq:plun_sum}. The bound weakens as $\rho(\lambda)$ grows, since more copies of $\lambda\cB$ are needed to cover $\cB$. When $\lambda\notin\Fp$, no nonzero $\bb$ is an integer multiple of $\lambda\bb$, and the code of Theorem~\ref{thm:additive}, whose bins are the shifted null spaces $\cA=\ba+\lambda U$ and $\cB=\bb+U$, has $|\cA+\lambda\cB|=|\cA|$ while $|\cA+\cB|=|\cA||\cB|$.

If $\lambda\notin\Fp$, Theorem~\ref{thm:additive} gives $\mathcal C=\mathcal C_{\mathrm{add}}=\cR_{\mathrm{cut}}$. If $\lambda\in\Fp$, the rate tuple $(1/2,1/2,1/2)\in\cR_{\mathrm{cut}}$ has sum-rate $3/2<3p/(2p-1)$, so by~\eqref{eq:plun_sum}, which holds for every $L$ and hence on the closure, it is not in $\mathcal C\supseteq\mathcal C_{\mathrm{add}}$. This gives the following criterion for cut-set optimality.

\begin{corollary}[Cut-set optimality]\label{cor:cutset}
For every $q$ and $\lambda\in\Fq\setminus\{0,1\}$, $\mathcal C=\cR_{\mathrm{cut}}$ if and only if $\lambda\notin\Fp$; and $\mathcal C_{\mathrm{add}}=\cR_{\mathrm{cut}}$ if and only if $\lambda\notin\Fp$.
\end{corollary}

\begin{remark}[The first open case]\label{rem:p7}
The first open case is $p=7$ with $\lambda\in\{3,5\}$. For $\lambda=3$, the six relabelings take only the values $3$ and $5$, and each user's pair in the proof of Theorem~\ref{thm:plunnecke} is $\{3,5\}$; since $\rho(3)=5$ and $\rho(5)=3$, the per-user inequalities~\eqref{eq:peruser} have $\rho_j=3$ for all three users, and adding $\sum_{i\neq j}R_i+R_j/3\geq4/3$ over $j$ gives $R_1+R_2+R_3\geq12/7$. The capacity therefore lies in $[12/7,2]$ for both cross-ratios.
\end{remark}

\subsection{Linear and additive codes}

For every cross-ratio, the linear capacity region is the decode-all region.

\begin{theorem}[Linear capacity region]\label{thm:linear}
For every $q$ and $\lambda\in\Fq\setminus\{0,1\}$, $\mathcal C_{\mathrm{lin}}=\cR_{\mathrm{all}}$.
\end{theorem}

The proof is in Section~\ref{sec:proof_linear}. The reason is clearest for one instance: an $\Fq$-linear message about one symbol is either a nonzero multiple of that symbol or nothing, so each user sends either its whole observation or none of it. No single observation determines the demand, so at least two users must send, and the server can then decode all the data.

For $L$ instances, the server's output is the sum of the contributions of the three users, each a matrix times that user's observation, and these must add up to $\ba+\lambda\bb$. Writing $\mathbf M$ for the contribution matrix of user~3, those of users~1 and~2 are forced to be $\mathbf I-\mathbf M$ and $\lambda\mathbf I-\mathbf M$. If $\mathbf M$ is diagonal, the scalar argument applies coordinate by coordinate: on a coordinate with entry $\theta$ the three users contribute $(1-\theta)a$, $(\lambda-\theta)b$ and $\theta s$, and since $0$, $1$ and $\lambda$ are distinct, at most one of $1-\theta$, $\lambda-\theta$ and $\theta$ is $0$, so at least two users send that coordinate. For a general $\mathbf M$, the proof uses the dimensions of its three eigenspaces instead.

The distinction from additive codes lies in the subspaces hidden by the encoders. An $\Fq$-linear encoder has an $\Fq$-subspace as its null space, which multiplication by $\lambda$ maps onto itself. Additive encoders can instead use $\Fp$-subspaces satisfying the separation condition $U\cap\lambda U=\{\boldsymbol 0\}$ of Section~\ref{sec:prime-free}.

When $\lambda\in\Fp$, identifying $\Fq$ with $\Fp^n$ turns an additive code of block length $L$ into an $\Fp$-linear code of block length $nL$ for the same cross-ratio and with the same rates, so Theorem~\ref{thm:linear} with $q=p$ gives the additive converse, and two users sending their observations achieve $\cR_{\mathrm{all}}$. Together with Theorem~\ref{thm:additive}, this gives the additive capacity region for every $\lambda$.

\begin{corollary}[Additive capacity region]\label{cor:additive}
For every $q$ and $\lambda\in\Fq\setminus\{0,1\}$,
\begin{equation*}
\mathcal C_{\mathrm{add}}=\begin{cases}\cR_{\mathrm{all}},&\lambda\in\Fp,\\ \cR_{\mathrm{cut}},&\lambda\notin\Fp.\end{cases}
\end{equation*}
\end{corollary}

\subsection{Consequences for rank, lattice data and entropy arguments}\label{sec:consequences}

All four-line instances have the same rank function and the same lattice data (Section~\ref{sec:fourline_model}), yet Theorems~\ref{thm:additive} and~\ref{thm:harmonic} give them different capacities within a single field.

\begin{corollary}[Capacity is not a function of the lattice data]\label{cor:lattice}
Over $\F_{3^2}$, the four-line instances with $\lambda=-1$ and with $\lambda\in\F_{3^2}\setminus\F_3$ have the same rank function, that of $U_{2,4}$, hence the same source entropy vector, and the same lattice data; the first has capacity $2$ and the second has capacity $3/2$. In the second, $\Fq$-linear codes have sum-rate at least $2$ while an additive code has sum-rate $3/2$.
\end{corollary}

\begin{remark}[Limits of entropic converses]\label{rem:entropic}
The corollary also limits converses based only on source entropies and functional dependencies. For the harmonic instance, such a converse would use the entropies of $\ba,\bb,\bs,\by,X_1,X_2,X_3$, together with $H(\ba)=H(\bb)=L$, $H(\ba,\bb)=2L$, the pairwise independence of $\ba,\bb,\bs,\by$ and the fact that any two determine the other two, the encoding constraints $H(X_1|\ba)=H(X_2|\bb)=H(X_3|\bs)=0$, and the decoding constraint $H(\by|X_1,X_2,X_3)=0$. These constraints do not distinguish the cross-ratios.

The additive code of Theorem~\ref{thm:additive} over $\F_{3^2}$ satisfies the same constraints with message entropies summing to $3L/2$. Hence no information inequality, Shannon or not, on these seven variables can derive $R_1+R_2+R_3\geq2$ from these constraints alone. An entropy converse for the harmonic bound needs additional constraints or auxiliary variables that encode the arithmetic of the cross-ratio; the proof in Section~\ref{sec:proof_harmonic} uses the code itself.
\end{remark}

\section{Proofs for the four-line instance}\label{sec:proofs}

\subsection{Proof of Theorem~\ref{thm:additive}}\label{sec:proof_additive}

We first work through the case $q=2^2$, $\lambda=\omega$, where $\F_{2^2}=\{0,1,\omega,\omega^2\}$ with $\omega^2=\omega+1$, so that $p=2$, $1-\lambda=1+\omega=\omega^2$ and $z\mapsto z^2$ swaps $\omega$ and $\omega^2$. The general case follows the same steps, with $\F_{2^2}$ replaced by $\Fq$ and $\omega$ by $\lambda$.

\begin{example}\label{ex:additive}
\emph{Frobenius subspace.} The subspace $U=\{(z,z^2)^\top:z\in\F_{2^2}\}$ and its multiples by $\omega$ and $\omega^2$ are
\begin{align*}
U&=\{(0,0)^\top,(1,1)^\top,(\omega,\omega^2)^\top,(\omega^2,\omega)^\top\},\\
\omega U&=\{(0,0)^\top,(\omega,\omega)^\top,(\omega^2,1)^\top,(1,\omega^2)^\top\},\\
\omega^2U&=\{(0,0)^\top,(\omega^2,\omega^2)^\top,(1,\omega)^\top,(\omega,1)^\top\},
\end{align*}
each closed under addition, since squaring preserves sums in characteristic $2$ (Example~\ref{ex:linadd}), and $U\cap\omega U=\{\boldsymbol 0\}$.

\emph{Messages.} With observations $\ba=(a_1,a_2)^\top$, $\bb=(b_1,b_2)^\top$ and $\bs=(s_1,s_2)^\top$, the users send
\[
X_1=a_2+\omega^2a_1^2,\qquad X_2=b_2+b_1^2,\qquad X_3=s_2+\omega s_1^2\ \in\F_{2^2}.
\]
These maps are additive, since squaring is $\F_2$-linear. For every $z\in\F_{2^2}$, using $\omega^3=1$, $\phi_1$ maps $(\omega z,\omega z^2)^\top\in\omega U$ to $\omega z^2+\omega^2(\omega z)^2=\omega z^2+\omega^4z^2=0$, $\phi_2$ maps $(z,z^2)^\top\in U$ to $z^2+z^2=0$, and $\phi_3$ maps $(\omega^2z,\omega^2z^2)^\top\in\omega^2U$ to $\omega^2z^2+\omega(\omega^2z)^2=\omega^2z^2+\omega^5z^2=0$. Each map takes all $4$ values, already when the first coordinate is $0$, so its null space has $16/4=4$ elements and is exactly $\omega U$, $U$ or $\omega^2U$, respectively. Each user sends one $4$-ary symbol per two instances: the rate tuple is $(1/2,1/2,1/2)$.

\emph{Decoding.} If two inputs have the same message triple, then, since $\phi_1$ is additive, it maps $\ba-\ba'$ to $0$, so $\ba-\ba'\in\omega U$, and in the same way $\bb-\bb'\in U$ and $\bs-\bs'\in\omega^2U$: $\ba-\ba'=\omega\bz_1$, $\bb-\bb'=\bz_2$ and $\bs-\bs'=\omega^2\bz_3$ with $\bz_1,\bz_2,\bz_3\in U$. Then $\bs-\bs'=(\ba-\ba')+(\bb-\bb')$ gives $\omega(\bz_1+\bz_3)=\bz_3+\bz_2\in\omega U\cap U=\{\boldsymbol 0\}$, so $\bz_1=\bz_2=\bz_3$ and $\by-\by'=\omega\bz_1+\omega\bz_2=\boldsymbol 0$. So all inputs with a given message triple have the same demand, and the server decodes by outputting this common demand.
\end{example}

\begin{proof}[Proof of Theorem~\ref{thm:additive}]
\emph{Frobenius subspace.} In characteristic $p$, raising to the power $p$ preserves sums and fixes the elements of $\Fp$, so the Frobenius map $z\mapsto z^p$ is $\Fp$-linear and
\begin{equation*}
U=\{(z,z^p)^\top:\ z\in\Fq\}\subseteq\Fq^2
\end{equation*}
is an $\Fp$-subspace with $|U|=q$, that is, of half the dimension of $\Fq^2$ over $\Fp$. The fixed points of $z\mapsto z^p$ are the roots of $z^p-z$. Every element of $\Fp$ is a root, and a polynomial of degree $p$ has at most $p$ roots, so the fixed points are exactly $\Fp$. If $\lambda(z,z^p)^\top\in U$, then $\lambda(z,z^p)^\top=(\bar z,\bar z^p)^\top$ for some $\bar z\in\Fq$; the first entries give $\bar z=\lambda z$, and the second entries give
\begin{equation*}
\lambda z^p=\lambda^p z^p,\qquad\text{so}\qquad(\lambda^p-\lambda)z^p=0,
\end{equation*}
and $\lambda^p\neq\lambda$ for $\lambda\notin\Fp$ gives $z=0$. Hence
\begin{equation}\label{eq:sep}
U\cap\lambda U=\{\boldsymbol 0\}.
\end{equation}

\emph{Messages.} With observations $\ba=(a_1,a_2)^\top$, $\bb=(b_1,b_2)^\top$ and $\bs=(s_1,s_2)^\top$, the users send
\begin{equation}\label{eq:addcode}
X_1=a_2-\lambda^{1-p}a_1^p,\qquad X_2=b_2-b_1^p,\qquad X_3=s_2-(1-\lambda)^{1-p}s_1^p\ \in\Fq.
\end{equation}
For $\gamma\neq 0$, consider the map $\bx\mapsto x_2-\gamma^{1-p}x_1^p$ on $\bx=(x_1,x_2)^\top\in\Fq^2$. It is additive, and it maps $\bx$ to $0$ exactly when $x_1$ determines $x_2$ as $x_2=\gamma^{1-p}x_1^p$, so its null space has $q$ elements. Every element $(\gamma z,\gamma z^p)^\top$ of $\gamma U$ lies in it, since $\gamma^{1-p}(\gamma z)^p=\gamma z^p$, and $\gamma U$ has $q$ elements, so the null space is $\gamma U$. With $\gamma=\lambda,1,1-\lambda$, the encoders~\eqref{eq:addcode} are additive, with null spaces $\lambda U$, $U$ and $(1-\lambda)U$. Each takes $q^2/q=q$ values, so each user sends one $q$-ary symbol per two instances.

\emph{Decoding.} Two observations of a user give the same message exactly when their difference lies in that user's null space, so if two inputs give the same message triple, there are $\bz_1,\bz_2,\bz_3\in U$ with
\begin{equation*}
\ba-\ba'=\lambda\bz_1,\qquad \bb-\bb'=\bz_2,\qquad \bs-\bs'=(1-\lambda)\bz_3.
\end{equation*}
By~\eqref{eq:c}, $\bs-\bs'=(\ba-\ba')+(\bb-\bb')$, so $\lambda\bz_1+\bz_2=\bz_3-\lambda\bz_3$ and
\begin{equation*}
\lambda(\bz_1+\bz_3)=\bz_3-\bz_2\ \in\ \lambda U\cap U.
\end{equation*}
By~\eqref{eq:sep}, $\bz_1=-\bz_3$ and $\bz_2=\bz_3$, so by~\eqref{eq:w}, $\by-\by'=(\ba-\ba')+\lambda(\bb-\bb')=-\lambda\bz_3+\lambda\bz_3=\boldsymbol 0$. The message triple determines the demand.

\emph{Region.} The vertex $(1/2,1/2,1/2)$ of $\cR_{\mathrm{cut}}$ is achieved, the other vertices are achieved by two users sending their observations, and time sharing and upward closure give $\cR_{\mathrm{cut}}\subseteq\mathcal C_{\mathrm{add}}$. With~\eqref{eq:cutset4}, $\mathcal C=\mathcal C_{\mathrm{add}}=\cR_{\mathrm{cut}}$.
\end{proof}

\subsection{Proof of Theorem~\ref{thm:harmonic}}\label{sec:proof_harmonic}

\begin{proofbody}
By the relabeling of Section~\ref{sec:fourline_model} it suffices to take $\lambda=-1$, so the server wants $\by=\ba-\bb$. Let $(\phi_1,\phi_2,\phi_3,\psi)$ be a zero-error code of block length $L$, and let two inputs $(\ba,\bb)$ and $(\ba',\bb')$ have the same message triple:
\begin{equation}\label{eq:same}
\phi_1(\ba)=\phi_1(\ba'),\qquad \phi_2(\bb)=\phi_2(\bb'),\qquad \phi_3(\ba+\bb)=\phi_3(\ba'+\bb').
\end{equation}

\emph{Exchange.} Zero error gives $\ba-\bb=\ba'-\bb'$, that is,
\begin{equation}\label{eq:d}
\ba-\ba'=\bb-\bb'.
\end{equation}
The exchanged inputs $(\ba,\bb')$ and $(\ba',\bb)$ satisfy $\phi_1(\ba)=\phi_1(\ba')$ and $\phi_2(\bb')=\phi_2(\bb)$ by~\eqref{eq:same}, and $\ba+\bb'=\ba'+\bb$ by~\eqref{eq:d}, so they have the same message triple.

\emph{Decoding.} Zero error applied to the exchanged inputs gives $\ba-\bb'=\ba'-\bb$, that is, $\ba-\ba'=\bb'-\bb$. With~\eqref{eq:d}, $2(\ba-\ba')=2(\bb-\bb')=\boldsymbol 0$, and as $q$ is odd, $\ba=\ba'$ and $\bb=\bb'$. Hence $(X_1,X_2,X_3)$ determines $(\ba,\bb)$.

\emph{Rate.} As $(X_1,X_2,X_3)$ determines $(\ba,\bb)$, the $q^{2L}$ inputs give distinct message triples, and $|\cX_1||\cX_2||\cX_3|\geq q^{2L}$. With $N_k=\log_q|\cX_k|$ and $R_k=N_k/L$, this is $N_1+N_2+N_3\geq 2L$, that is, $R_1+R_2+R_3\geq 2$. With~\eqref{eq:cutset4}, $\mathcal C\subseteq\cR_{\mathrm{all}}$, and two users sending their observations achieve the vertices of $\cR_{\mathrm{all}}$, so $\cR_{\mathrm{all}}\subseteq\mathcal C$.
\end{proofbody}

\subsection{Proof of Theorem~\ref{thm:plunnecke}}\label{sec:proof_plunnecke}

\begin{proofbody}
Let $(\phi_1,\phi_2,\phi_3,\psi)$ be a zero-error code of block length $L$, and let $\cA,\cB\subseteq\Fq^L$ be largest bins of $\phi_1$ and $\phi_2$. The $|\cX_1|$ bins of $\phi_1$ partition the $q^L$ observations of user~1, so by the pigeonhole principle the largest has at least $q^L/|\cX_1|$ of them, and similarly for $\phi_2$:
\begin{equation}\label{eq:bins}
|\cA|\geq q^L/|\cX_1|,\qquad |\cB|\geq q^L/|\cX_2|.
\end{equation}

\emph{Distinct sums.} If $\ba+\bb=\ba'+\bb'$ with $\ba,\ba'\in\cA$ and $\bb,\bb'\in\cB$, then $(\ba,\bb)$ and $(\ba',\bb')$ have the same message triple, so $\ba+\lambda\bb=\ba'+\lambda\bb'$ by zero error, and subtracting, $(\lambda-1)(\bb-\bb')=\boldsymbol 0$, so $\bb=\bb'$ and $\ba=\ba'$. Thus distinct pairs $(\ba,\bb)\in\cA\times\cB$ have distinct sums, so the set $\cA+\cB$ of sums has as many elements as $\cA\times\cB$; the same holds with $\cA$ replaced by any nonempty $\cA'\subseteq\cA$:
\begin{equation}\label{eq:inj}
|\cA'+\cB|=|\cA'||\cB|.
\end{equation}

\emph{Demand image.} With the messages of users 1 and 2 fixed to those of $\cA$ and $\cB$, the demand is a function of the third message, so
\begin{equation}\label{eq:image}
|\cA+\lambda\cB|\leq|\cX_3|.
\end{equation}

\emph{Pl\"unnecke--Ruzsa.} Let
\[
\tau=\frac{|\cA+\lambda\cB|}{|\cA|},\qquad\text{where }\lambda\cB=\{\lambda\bb:\bb\in\cB\},
\]
so that $\tau$ is the factor by which adding $\lambda\cB$ enlarges $\cA$. The Pl\"unnecke--Ruzsa inequality, in the form of~\cite{Petridis}, valid in every abelian group, bounds repeated sums by this factor: there is a nonempty $\cA'\subseteq\cA$ with
\begin{equation}\label{eq:petridis}
|\cA'+\underbrace{\lambda\cB+\cdots+\lambda\cB}_{\rho(\lambda)}|\leq \tau^{\rho(\lambda)}|\cA'|.
\end{equation}
Since $\rho(\lambda)\lambda=1$ in $\Fp$, every $\bb\in\cB$ equals $\lambda\bb+\cdots+\lambda\bb$ with $\rho(\lambda)$ terms, each in $\lambda\cB$. Hence $\cB\subseteq\lambda\cB+\cdots+\lambda\cB$ ($\rho(\lambda)$ terms) and $\cA'+\cB\subseteq\cA'+\lambda\cB+\cdots+\lambda\cB$. With~\eqref{eq:inj} and~\eqref{eq:petridis},
\begin{equation*}
|\cA'||\cB|=|\cA'+\cB|\leq|\cA'+\lambda\cB+\cdots+\lambda\cB|\leq \tau^{\rho(\lambda)}|\cA'|.
\end{equation*}
Dividing by $|\cA'|$ gives $\tau\geq|\cB|^{1/\rho(\lambda)}$, that is,
\begin{equation}\label{eq:sumset_bound}
|\cA+\lambda\cB|=\tau|\cA|\geq|\cA||\cB|^{1/\rho(\lambda)}.
\end{equation}

\emph{Rate.} By~\eqref{eq:image}, \eqref{eq:sumset_bound} and~\eqref{eq:bins},
\begin{equation*}
|\cX_3|\geq|\cA+\lambda\cB|\geq|\cA||\cB|^{1/\rho(\lambda)}\geq\frac{q^L}{|\cX_1|}\Big(\frac{q^L}{|\cX_2|}\Big)^{1/\rho(\lambda)}.
\end{equation*}
Taking $\log_q$, with $N_k=\log_q|\cX_k|$, gives $N_3\geq(L-N_1)+(L-N_2)/\rho(\lambda)$, that is, $N_1+N_3+N_2/\rho(\lambda)\geq L(1+1/\rho(\lambda))$. Dividing by $L$ and using $R_k=N_k/L$ gives
\begin{equation}\label{eq:plun}
R_1+R_3+\frac{R_2}{\rho(\lambda)}\ \geq\ 1+\frac{1}{\rho(\lambda)}.
\end{equation}

\emph{Relabeling.} Relabeling the users can put any user $j$ in the role of the user observing $\bb$, with the cross-ratio drawn from a pair of relabelings of $\lambda$. The pairs are $\{\lambda,1-\lambda\}$ for user~2, $\{1/\lambda,(\lambda-1)/\lambda\}$ for user~1, and $\{1/(1-\lambda),\lambda/(\lambda-1)\}$ for user~3. The two members of a pair correspond to the two ways of assigning the other two users to the roles of the users observing $\ba$ and $\bs$. For user~2, keeping the labels gives $\lambda$, and exchanging users~1 and~3 with $\bb$ replaced by $-\bb$ gives $1-\lambda$, since $\ba+\lambda\bb=\bs+(1-\lambda)(-\bb)$; the pairs of users~1 and~3 follow in the same way. Write $\rho_j$ for the smaller value of $\rho$ on user $j$'s pair. Both members of each pair lie in $\Fp\setminus\{0,1\}$, since the relabelings of $\lambda$ stay in $\Fp$. Applying~\eqref{eq:plun} to the labeling with the smaller $\rho$ gives, for each $j$,
\begin{equation}\label{eq:peruser}
\sum_{i\neq j}R_i+\frac{R_j}{\rho_j}\ \geq\ 1+\frac{1}{\rho_j}.
\end{equation}
Write~\eqref{eq:peruser} as $\sum_{i\neq j}R_i-1\geq(1-R_j)/\rho_j$. Since $\rho_j\leq p-1$, the right side is at least $(1-R_j)/(p-1)$ if $R_j\leq1$, and since $\sum_{i\neq j}R_i\geq 1$ by~\eqref{eq:cutset4}, the left side is at least $(1-R_j)/(p-1)$ if $R_j>1$. Hence $\sum_{i\neq j}R_i+R_j/(p-1)\geq 1+1/(p-1)$ for each $j$; adding the three inequalities, $\frac{2p-1}{p-1}(R_1+R_2+R_3)\geq\frac{3p}{p-1}$, which is~\eqref{eq:plun_sum}.
\end{proofbody}

\subsection{Proof of Theorem~\ref{thm:linear}}\label{sec:proof_linear}

\begin{proofbody}
\emph{Code.} Let an $\Fq$-linear zero-error code have block length $L$. Each user observes one vector in $\Fq^L$, namely $\ba$, $\bb$ or $\bs$, so user $k$ sends $X_k=\mathbf F_k\ba$, $\mathbf F_k\bb$ or $\mathbf F_k\bs$ with $\mathbf F_k\in\Fq^{N_k\times L}$, and $R_k=N_k/L$. By Footnote~\ref{fn:linear-decoder}, the decoder can be taken linear: $\by=\mathbf U_1X_1+\mathbf U_2X_2+\mathbf U_3X_3$ with $\mathbf U_k\in\Fq^{L\times N_k}$.

\emph{Reduction to one matrix.} Substituting $\bs=\ba+\bb$,
\begin{equation*}
\ba+\lambda\bb=\by=(\mathbf U_1\mathbf F_1+\mathbf U_3\mathbf F_3)\ba+(\mathbf U_2\mathbf F_2+\mathbf U_3\mathbf F_3)\bb\qquad\text{for all }\ba,\bb\in\Fq^L.
\end{equation*}
Since this equality holds for all $\ba$ and $\bb$, the matrices multiplying $\ba$ on the two sides are equal, and so are those multiplying $\bb$: $\mathbf U_1\mathbf F_1+\mathbf U_3\mathbf F_3=\mathbf I$ and $\mathbf U_2\mathbf F_2+\mathbf U_3\mathbf F_3=\lambda\mathbf I$. With the $L\times L$ matrix $\mathbf M=\mathbf U_3\mathbf F_3$, these equalities become
\begin{equation}\label{eq:linP}
\mathbf U_1\mathbf F_1=\mathbf I-\mathbf M,\qquad \mathbf U_2\mathbf F_2=\lambda\mathbf I-\mathbf M,\qquad \mathbf U_3\mathbf F_3=\mathbf M.
\end{equation}
Since $\rank(\mathbf U_k\mathbf F_k)\leq\rank\mathbf F_k\leq N_k$, the three matrices in~\eqref{eq:linP} have ranks at most $N_1$, $N_2$ and $N_3$.

\emph{Rank inequality.} For every $\mathbf M\in\Fq^{L\times L}$,
\begin{equation}\label{eq:eig}
\rank\mathbf M+\rank(\mathbf I-\mathbf M)+\rank(\lambda\mathbf I-\mathbf M)\geq 2L.
\end{equation}
Indeed, for $\theta\in\{0,1,\lambda\}$, $\nullspace(\theta\mathbf I-\mathbf M)=\{\bx\in\Fq^L:\mathbf M\bx=\theta\bx\}$, and $\rank(\theta\mathbf I-\mathbf M)=L-\dim\nullspace(\theta\mathbf I-\mathbf M)$. If $\bx_0+\bx_1+\bx_\lambda=\boldsymbol 0$ with $\bx_\theta\in\nullspace(\theta\mathbf I-\mathbf M)$, multiplying by $\mathbf M$ once and twice gives $\bx_1+\lambda\bx_\lambda=\boldsymbol 0$ and $\bx_1+\lambda^2\bx_\lambda=\boldsymbol 0$, so $\lambda(\lambda-1)\bx_\lambda=\boldsymbol 0$; as $\lambda\notin\{0,1\}$, $\bx_\lambda=\boldsymbol 0$, and then $\bx_1=\bx_0=\boldsymbol 0$. Now take a basis of each of the three null spaces. A vanishing linear combination of all these vectors splits as $\bx_0+\bx_1+\bx_\lambda=\boldsymbol 0$ with $\bx_\theta\in\nullspace(\theta\mathbf I-\mathbf M)$, so each $\bx_\theta=\boldsymbol 0$ and all coefficients vanish. So the union of the three bases is linearly independent in $\Fq^L$, so it has at most $L$ vectors: the three dimensions add to at most $L$, and
\begin{equation*}
\rank\mathbf M+\rank(\mathbf I-\mathbf M)+\rank(\lambda\mathbf I-\mathbf M)=3L-\sum_{\theta\in\{0,1,\lambda\}}\dim\nullspace(\theta\mathbf I-\mathbf M)\geq 3L-L=2L.
\end{equation*}

\emph{Converse.} By~\eqref{eq:linP} and~\eqref{eq:eig}, $N_1+N_2+N_3\geq 2L$, that is, $R_1+R_2+R_3\geq 2$ for every $\Fq$-linear zero-error code. With the cut-set bounds~\eqref{eq:cutset4}, every such rate tuple lies in $\cR_{\mathrm{all}}$, and so does the closure: $\mathcal C_{\mathrm{lin}}\subseteq\cR_{\mathrm{all}}$.

\emph{Achievability.} The vertices $(1,1,0)$, $(1,0,1)$ and $(0,1,1)$ of $\cR_{\mathrm{all}}$ are achieved with $L=1$ by sending two of the observations, which determine $\by$; the rest of $\cR_{\mathrm{all}}$ follows by time sharing and upward closure. In terms of~\eqref{eq:linP} these are $\mathbf M=\mathbf 0$, $\mathbf M=\lambda\mathbf I$ and $\mathbf M=\mathbf I$.
\end{proofbody}

\section{Discussion}\label{sec:discussion}

For non-harmonic $\lambda\in\Fp$, which requires $p\geq7$, the capacity lies in $[3p/(2p-1),2]$, and either bound may be improved; the smallest case is $p=7$, $\lambda\in\{3,5\}$, with $[12/7,2]$. Since the coefficients $0,1,\lambda$ lie in $\Fp$, viewing each symbol of $\Fq$ as $n$ symbols of $\Fp$ shows that the capacity region over $\Fq$ equals that over $\Fp$, so these are problems over prime fields. We outline possible directions.

\emph{Better converses.} Any converse above $3/2$ must use that $\lambda\in\Fp$, since the additive code of Theorem~\ref{thm:additive} satisfies the same entropic constraints (Remark~\ref{rem:entropic}); Theorems~\ref{thm:plunnecke} and~\ref{thm:ve} use it through $\rho(\lambda)$ and $\nu(\lambda)$. The sumset bound uses only the bins of two users and the weakest form of the Pl\"unnecke--Ruzsa inequality, and it is not tight for some $\lambda$: the vanishing-error bound~\eqref{eq:ve_prime} also holds with zero error and is larger when $\nu(\lambda)$ is small compared with $p$ (Remark~\ref{rem:ve_compare}). Using the bins of all three users at once, or combining $\rho(\lambda)$ and $\nu(\lambda)$, may improve both.

\emph{Better codes.} Additive codes need $2$ for $\lambda\in\Fp$, so a code below $2$ must be nonadditive. It needs bins $\cA$ and $\cB$ of users~1 and~2 with $(\cA-\cA)\cap\lambda(\cB-\cB)$ large, so that many pairs share a demand, and $(\cA-\cA)\cap(\cB-\cB)$ trivial, as zero error requires. $\Fp$-subspaces do this only when $\lambda\notin\Fp$ (Theorem~\ref{thm:additive}), so other structures are needed, such as the intervals of integers used in~\cite{Con_Tamo}; which structures are best, and how close any of them can come to the cut-set bound, is not known.

\emph{A test case.} A natural place to look in both directions is an \emph{equianharmonic} $\lambda$, a root of $\lambda^2-\lambda+1=0$, such as $\lambda\in\{3,5\}$ over $\F_7$; for $p=3$ the only root is $\lambda=-1$, which is harmonic. Its six relabelings take only the values $\lambda$ and $1-\lambda$, and with $\widetilde{\ba}=-\bb$ and $\widetilde{\bb}=\ba+\bb$ the observations become $(-\bb,\bs,\ba)$ while the demand becomes $\widetilde{\ba}+\lambda\widetilde{\bb}=\lambda(\ba+\lambda\bb)$, since $\lambda^2=\lambda-1$. So a code with rate tuple $(R_1,R_2,R_3)$ gives one with rate tuple $(R_3,R_1,R_2)$: the problem is symmetric under cyclic permutations of the users, and codes with this symmetry are natural candidates to improve on $2$, while a converse may exploit it.

\emph{Vanishing error.} For uniform data, Appendix~\ref{app:ve} settles $\lambda\notin\Fp$, where the vanishing-error capacity is $3/2$, and leaves every $\lambda\in\Fp$ open: the vanishing-error capacity lies in $[12/7,2]$ for harmonic $\lambda$ and in $[3(3\nu-2)/(6\nu-5),2]$ for the other $\lambda\in\Fp$, with $\nu=\nu(\lambda)$ as in Theorem~\ref{thm:ve}. The sharpest question is the harmonic one: whether vanishing error forces $H(\ba,\bb\,|\,X_1,X_2,X_3)=o(L)$, which would give capacity $2$ under both criteria. More broadly, for uniform data the four-line instance with $\lambda\in\Fp$ is the smallest configuration for which it is not known whether the zero-error and vanishing-error capacities coincide.

\appendix

\section{Vanishing error}\label{app:ve}

The separation between cross-ratios inside and outside the prime subfield persists under vanishing error for uniform data. We first define the capacity regions and state the results, then develop the entropy inequalities used in the converses and prove the bounds. The final subsection lists the zero-error results that do not carry over.

\subsection{Vanishing-error capacity and the results}

We retain the uniform data, encoders and decoder of Section~\ref{sec:model}, and replace exact recovery for every input by a decoding error probability that tends to zero. For a code of block length $L$, define
\begin{equation*}
\epsilon_L=\Pr\big(\psi(X_1,\ldots,X_K)\neq\mathbf D^{[L]}\bw^L\big).
\end{equation*}
A rate tuple $(R_1,\ldots,R_K)$ is \emph{achievable with vanishing error} if there is a sequence of codes with block lengths $L\to\infty$, rates at most $R_1,\ldots,R_K$, and $\epsilon_L\to 0$. The \emph{vanishing-error capacity region} $\mathcal C^{\mathrm{ve}}$ is the closure of the set of these tuples. The \emph{vanishing-error capacity} is the minimum of $R_1+\cdots+R_K$ over this region. Restricting the encoders to $\Fq$-linear maps or to additive ($\Fp$-linear) maps, while allowing an arbitrary decoder, defines $\mathcal C^{\mathrm{ve}}_{\mathrm{lin}}$ and $\mathcal C^{\mathrm{ve}}_{\mathrm{add}}$, respectively.

Repeating any zero-error code on independent blocks gives arbitrarily large block lengths with the same rates and zero error, so $\mathcal C\subseteq\mathcal C^{\mathrm{ve}}$. Consequently, every vanishing-error converse below also applies to zero-error codes. The changes of basis and relabelings in Section~\ref{sec:model} preserve the uniform input distribution and the decoding error probability, so the corresponding invariance statements also hold for $\mathcal C^{\mathrm{ve}}$.

Recall that the relabelings of $\lambda$ are the six values $\lambda$, $1-\lambda$, $1/\lambda$, $1/(1-\lambda)$, $(\lambda-1)/\lambda$ and $\lambda/(\lambda-1)$ (Section~\ref{sec:fourline_model}), and identify $\Fp$ with $\{0,1,\ldots,p-1\}$ as in Section~\ref{sec:prime}. For $\lambda\in\Fp\setminus\{0,1\}$ let $\nu(\lambda)$ be the smallest relabeling of $\lambda$. The relabelings avoid $0$ and $1$, so $\nu(\lambda)\geq2$. As integers, the relabeling $1-\lambda$ equals $p+1-\lambda$, and $\min\{\lambda,p+1-\lambda\}\leq(p+1)/2$, so $\nu(\lambda)\leq(p+1)/2$. Moreover, $\nu(\lambda)=2$ exactly when $\lambda$ is harmonic, since the six relabelings of $2$ are $2$, $1-2=-1$, $1/2$, $1/(1-2)=-1$, $(2-1)/2=1/2$ and $2/(2-1)=2$, which are the harmonic values. The size of $\nu(\lambda)$ as an integer matters because the proof uses entropy inequalities for repeated addition. In Theorem~\ref{thm:plunnecke} the integer $\rho(\lambda)=1/\lambda$ plays a similar role.

\begin{theorem}[Vanishing error]\label{thm:ve}
For the four-line instance with $\lambda\in\Fq\setminus\{0,1\}$:
\begin{enumerate}
\item If $\lambda\notin\Fp$, then $\mathcal C^{\mathrm{ve}}=\mathcal C=\cR_{\mathrm{cut}}$, and the vanishing-error capacity is $3/2$.
\item If $\lambda\in\Fp$, then every $(R_1,R_2,R_3)\in\mathcal C^{\mathrm{ve}}$ satisfies, with $\nu=\nu(\lambda)$,
\begin{equation}\label{eq:ve_prime}
R_1+R_2+R_3\ \geq\ \frac{3(3\nu-2)}{6\nu-5}\ \geq\ \frac{3(3p-1)}{6p-4}\ >\ \frac32.
\end{equation}
Hence $\mathcal C^{\mathrm{ve}}=\cR_{\mathrm{cut}}$ if and only if $\lambda\notin\Fp$, and for harmonic $\lambda$, where $\nu=2$, the vanishing-error capacity lies in $[12/7,2]$.
\item For every $\lambda$, $\mathcal C^{\mathrm{ve}}_{\mathrm{lin}}=\mathcal C_{\mathrm{lin}}=\cR_{\mathrm{all}}$ and $\mathcal C^{\mathrm{ve}}_{\mathrm{add}}=\mathcal C_{\mathrm{add}}$.
\end{enumerate}
\end{theorem}

\begin{remark}[Comparison with zero error]\label{rem:ve_compare}
Parts 1 and 2 show that the vanishing-error capacity region is the cut-set region exactly when $\lambda\notin\Fp$, as with zero error, and part 2 keeps a gap above $3/2$ for every $\lambda\in\Fp$. The harmonic bound $12/7$ is weaker than the zero-error value $2$. For the other $\lambda\in\Fp$, the weighted zero-error bounds~\eqref{eq:peruser} from the proof of Theorem~\ref{thm:plunnecke} and the bound~\eqref{eq:ve_prime} are incomparable, and either can be the stronger: for $p=7$ and $\lambda\in\{3,5\}$, where $\nu=3$, the bound~\eqref{eq:ve_prime} is $21/13\approx 1.6$, smaller than $12/7\approx 1.7$ from~\eqref{eq:peruser}, while for $p=101$ and $\lambda=3$, where $(\rho_1,\rho_2,\rho_3)=(3,34,68)$, the weighted bounds~\eqref{eq:peruser} with the cut-set bounds allow sum-rate $18088/11255\approx 1.6071$ and~\eqref{eq:ve_prime} requires $21/13\approx 1.6154$. Since $\mathcal C\subseteq\mathcal C^{\mathrm{ve}}$, \eqref{eq:ve_prime} is also a zero-error converse, and for some $\lambda$, such as $p=101$ and $\lambda=3$, it strengthens Theorem~\ref{thm:plunnecke}. The equalities $\mathcal C^{\mathrm{ve}}_{\mathrm{lin}}=\mathcal C_{\mathrm{lin}}$ and $\mathcal C^{\mathrm{ve}}_{\mathrm{add}}=\mathcal C_{\mathrm{add}}$ hold for the general problem.
\end{remark}

The two-user capacity region is also unchanged under vanishing error.

\begin{proposition}[Two users with vanishing error]\label{prop:ve_twouser}
For $K=2$, $\mathcal C^{\mathrm{ve}}=\mathcal C$, the region of Theorem~\ref{thm:twouser}.
\end{proposition}

With Theorem~\ref{thm:ve}, Corollary~\ref{cor:lattice} holds with vanishing error: over $\F_{3^2}$ the four-line instances with $\lambda=-1$ and with $\lambda\notin\F_3$ have the same rank function and the same lattice data, and their vanishing-error capacities are at least $12/7$ and equal to $3/2$. Table~\ref{tab:ve} compares the two criteria.

\begin{table}[H]
\centering
\renewcommand{\arraystretch}{1.15}
\begin{tabular}{lll}
\toprule
& zero error & vanishing error\\
\midrule
$\lambda\notin\Fp$ & $\mathcal C=\cR_{\mathrm{cut}}$, capacity $3/2$ & $\mathcal C^{\mathrm{ve}}=\cR_{\mathrm{cut}}$, capacity $3/2$\\
$\lambda$ harmonic, $q$ odd & $\mathcal C=\cR_{\mathrm{all}}$, capacity $2$ & capacity in $[12/7,2]$\\
$\lambda\in\Fp$, not harmonic & capacity in $[3p/(2p-1),2]$ & capacity in $[3(3\nu-2)/(6\nu-5),2]$\\
$\Fq$-linear and additive codes & $\mathcal C_{\mathrm{lin}}$, $\mathcal C_{\mathrm{add}}$ & the same regions\\
two users & Theorem~\ref{thm:twouser} & the same region\\
\bottomrule
\end{tabular}
\caption{Zero error and vanishing error for uniform data, with $\nu=\nu(\lambda)$.}\label{tab:ve}
\end{table}

\subsection{Entropy of sums}

The zero-error converses use cardinalities of sumsets of bins. For vanishing error, we use entropies of sums of independent random vectors. The required inequalities are entropy analogues of sumset inequalities, in the spirit of~\cite{Tao_entropy}; we prove them here to keep the argument self-contained, using $\bxi_1,\bxi_2,\ldots$ for generic random vectors in $\Fq^L$.

First, if $\bxi_1,\bxi_2,\bxi_3$ are independent, then
\begin{equation}\label{eq:submod}
H(\bxi_1+\bxi_2+\bxi_3)\leq H(\bxi_1+\bxi_2)+H(\bxi_2+\bxi_3)-H(\bxi_2).
\end{equation}
Indeed,
\begin{equation*}
H(\bxi_1+\bxi_2+\bxi_3)-H(\bxi_2+\bxi_3)=I(\bxi_1;\bxi_1+\bxi_2+\bxi_3)\leq I(\bxi_1;\bxi_1+\bxi_2)=H(\bxi_1+\bxi_2)-H(\bxi_2),
\end{equation*}
where the equalities hold because $\bxi_1$ is independent of $(\bxi_2,\bxi_3)$, and the inequality is the data processing inequality for the Markov chain $\bxi_1-(\bxi_1+\bxi_2)-(\bxi_1+\bxi_2+\bxi_3)$, which holds because $\bxi_3$ is independent of $(\bxi_1,\bxi_2)$.

Second, if $\bxi_3$ is independent of the pair $(\bxi_1,\bxi_2)$, with no independence assumption within that pair, then
\begin{equation}\label{eq:triangle}
H(\bxi_1-\bxi_2)+H(\bxi_3)\leq H(\bxi_1-\bxi_3)+H(\bxi_3-\bxi_2).
\end{equation}
To prove this, the pair $(\bxi_1-\bxi_3,\bxi_3-\bxi_2)$ determines $\bxi_1-\bxi_2$, so
\begin{align*}
H(\bxi_1-\bxi_3)+H(\bxi_3-\bxi_2)
&\geq H(\bxi_1-\bxi_3,\bxi_3-\bxi_2)\\
&=H(\bxi_1-\bxi_2)+H(\bxi_1-\bxi_3\,|\,\bxi_1-\bxi_2)\\
&\geq H(\bxi_1-\bxi_2)+H(\bxi_3\,|\,\bxi_1,\bxi_2)\\
&=H(\bxi_1-\bxi_2)+H(\bxi_3).
\end{align*}
For a random vector $\bxi$ with an independent copy $\bxi'$, let
\begin{equation*}
g(\bxi)=H(\bxi-\bxi')-H(\bxi)\geq 0
\end{equation*}
be the \emph{growth} of $\bxi$, the entropy added by subtracting an independent copy. The following inequality, built from~\eqref{eq:submod} and~\eqref{eq:triangle}, plays the role of the Pl\"unnecke--Ruzsa inequality in the proof of Theorem~\ref{thm:plunnecke}. It bounds the increase from $H(\bxi_1-\bxi_2)$ to $H(\bxi_1+r\bxi_2)$: each of $r-1$ additions costs at most the bracket below, and one reflection costs at most the smaller of the two values of $g$.

\begin{lemma}[Entropic sumset inequality]\label{lem:entropy}
Let $\bxi_1$ and $\bxi_2$ be independent random vectors in $\Fq^L$, and let $\bxi_1'$ and $\bxi_2'$ have the distributions of $\bxi_1$ and $\bxi_2$, respectively, with $\bxi_1,\bxi_2,\bxi_1',\bxi_2'$ independent. For every positive integer $r$ that is not a multiple of $p$,
\begin{equation}\label{eq:key}
H(\bxi_1+r\bxi_2)-H(\bxi_1-\bxi_2)\leq(r-1)\big[H(\bxi_1-\bxi_2)-H(\bxi_1)+g(\bxi_1)\big]+\min\{g(\bxi_1),g(\bxi_2)\}.
\end{equation}
\end{lemma}

\begin{proof}
\emph{Addition.} For an integer $m$, apply~\eqref{eq:triangle} to the three vectors $\bxi_1-m\bxi_2$, $\bxi_2$ and $\bxi_1'$. Since $(\bxi_1',\bxi_2)$ has the same distribution as $(\bxi_1,\bxi_2)$,
\begin{align*}
H(\bxi_1-(m+1)\bxi_2)+H(\bxi_1)
&\leq H(\bxi_1-\bxi_1'-m\bxi_2)+H(\bxi_1-\bxi_2)\\
&\leq H(\bxi_1-\bxi_1')+H(\bxi_1-m\bxi_2)-H(\bxi_1)+H(\bxi_1-\bxi_2),
\end{align*}
where the second inequality is~\eqref{eq:submod} applied to $-\bxi_1'$, $\bxi_1$ and $-m\bxi_2$. That is,
\begin{equation*}
H(\bxi_1-(m+1)\bxi_2)-H(\bxi_1-m\bxi_2)\leq H(\bxi_1-\bxi_2)-H(\bxi_1)+g(\bxi_1),
\end{equation*}
and adding these inequalities for $m=1,\ldots,r-1$ bounds $H(\bxi_1-r\bxi_2)-H(\bxi_1-\bxi_2)$ by the first term on the right of~\eqref{eq:key}.

\emph{Reflection.} For every integer $m$, adding the independent vector $-\bxi_1'$ does not decrease entropy, and~\eqref{eq:submod} applied to $\bxi_1$, $-\bxi_1'$ and $m\bxi_2$, with $(\bxi_1',\bxi_2)$ distributed as $(\bxi_1,\bxi_2)$, gives
\begin{align}
H(\bxi_1+m\bxi_2)&\leq H(\bxi_1-\bxi_1'+m\bxi_2)\nonumber\\
&\leq H(\bxi_1-m\bxi_2)+g(\bxi_1).\label{eq:refl1}
\end{align}
Applying~\eqref{eq:refl1} to $m\bxi_2$ and $\bxi_1$ in place of $\bxi_1$ and $\bxi_2$, with $m=1$, and using that multiplication by the integer $m$, that is, $m$-fold addition, is invertible on $\Fq^L$ and preserves entropy when $m$ is not a multiple of $p$,
\begin{equation}\label{eq:refl2}
H(\bxi_1+m\bxi_2)\leq H(\bxi_1-m\bxi_2)+g(\bxi_2).
\end{equation}
With $m=r$, \eqref{eq:refl1} and~\eqref{eq:refl2} bound $H(\bxi_1+r\bxi_2)-H(\bxi_1-r\bxi_2)$ by $\min\{g(\bxi_1),g(\bxi_2)\}$, and adding the addition bound gives~\eqref{eq:key}.
\end{proof}

\subsection{Fano's inequality and conditional entropies}

For the four-line instance, let $X=(X_1,X_2,X_3)$ and $\delta_L=H(\by\,|\,X)$. By Fano's inequality, $\delta_L/L\to 0$ when $\epsilon_L\to 0$. Write $\boldsymbol o_1=\ba$, $\boldsymbol o_2=\bb$ and $\boldsymbol o_3=\bs$ for the observations of users $1$, $2$ and $3$, so that one statement covers all users. Given $X_i$, generate $\boldsymbol o_i'$ from the conditional distribution of $\boldsymbol o_i$, using fresh randomness independent of the data and of the other users' copies, and let
\begin{equation*}
g_i=H(\boldsymbol o_i-\boldsymbol o_i'\,|\,X_i)-H(\boldsymbol o_i\,|\,X_i),
\end{equation*}
the entropy added by subtracting an independent observation from the same bin of user $i$. For distinct $i,j,k$, let
\begin{equation*}
\sigma_i=H(X_j)+H(X_k)-L+\delta_L,\qquad \sigma=\sigma_1+\sigma_2+\sigma_3=2\big(H(X_1)+H(X_2)+H(X_3)\big)-3(L-\delta_L).
\end{equation*}
These are the \emph{cut slacks}: the demand is uniform and independent of $\boldsymbol o_i$, so $L=H(\by\,|\,\boldsymbol o_i)\leq H(X_j)+H(X_k)+\delta_L$ and $\sigma_i\geq0$, and $\sigma$ measures the communication above the cut-set bound.

\begin{lemma}[Conditional entropies]\label{lem:cells}
Every code of block length $L$ satisfies, for distinct $i,j,k$, the Markov chain~\eqref{eq:cell_markov}, the conditional entropy bounds~\eqref{eq:cell_bin}--\eqref{eq:cell_sum} and the difference bound~\eqref{eq:cell_diff}:
\begin{align}
\boldsymbol o_i\ -\ X_i\ &-\ (X_j,\boldsymbol o_j)\quad\text{is a Markov chain},\label{eq:cell_markov}\\
H(\boldsymbol o_i\,|\,X_i)&=L-H(X_i),\label{eq:cell_bin}\\
H(\by\,|\,X_i,X_j)&\leq H(X_k)+\delta_L,\label{eq:cell_dem}\\
H(\boldsymbol o_k\,|\,X_i,X_j)&\geq 2L-H(X_i)-H(X_j)-\delta_L,\label{eq:cell_sum}\\
0\leq g_i&\leq \sigma_i+\min\{\sigma_j,\sigma_k\}.\label{eq:cell_diff}
\end{align}
\end{lemma}

\begin{proof}
\emph{Markov chain.} The pairs $(\boldsymbol o_i,X_i)$ and $(\boldsymbol o_j,X_j)$ are independent, which gives~\eqref{eq:cell_markov}. In particular, given $(X_i,X_j)$, the observations $\boldsymbol o_i$ and $\boldsymbol o_j$ are independent, and $\boldsymbol o_i$ has the same conditional distribution as given $X_i$.

\emph{Conditional entropies.} Since $X_i$ is a function of $\boldsymbol o_i$, $H(\boldsymbol o_i\,|\,X_i)=H(\boldsymbol o_i)-H(X_i)=L-H(X_i)$, which is~\eqref{eq:cell_bin}. For~\eqref{eq:cell_dem},
\begin{equation*}
H(\by\,|\,X_i,X_j)\leq H(X_k,\by\,|\,X_i,X_j)=H(X_k\,|\,X_i,X_j)+H(\by\,|\,X)\leq H(X_k)+\delta_L.
\end{equation*}
For~\eqref{eq:cell_sum},
\begin{align*}
H(\boldsymbol o_k\,|\,X_i,X_j)&\geq H(\boldsymbol o_i,\boldsymbol o_j\,|\,X_i,X_j)-H(\boldsymbol o_i,\boldsymbol o_j\,|\,X,\boldsymbol o_k)\\
&=H(\boldsymbol o_i,\boldsymbol o_j)-H(X_i,X_j)-H(\by\,|\,X,\boldsymbol o_k)\\
&\geq 2L-H(X_i)-H(X_j)-\delta_L,
\end{align*}
where the first step is $H(\boldsymbol o_k\,|\,X_i,X_j)\geq I(\boldsymbol o_k;\boldsymbol o_i,\boldsymbol o_j\,|\,X_i,X_j)$ with $X_k$ a function of $\boldsymbol o_k$, and the equality uses that $X_i$ and $X_j$ are functions of $(\boldsymbol o_i,\boldsymbol o_j)$ and that, given $\boldsymbol o_k$, the demand and $(\boldsymbol o_i,\boldsymbol o_j)$ determine each other.

\emph{Difference bound.} Given $X_i$, $\boldsymbol o_i'$ is independent of $\boldsymbol o_i$, so $H(\boldsymbol o_i-\boldsymbol o_i'\,|\,X_i)\geq H(\boldsymbol o_i\,|\,X_i)$ and $g_i\geq0$. Any two observations determine the demand, so $\by=\alpha\boldsymbol o_i+\beta\boldsymbol o_j$ with $\alpha,\beta\neq 0$. Given $(X_i,X_j)$,
\begin{align*}
g_i+2L-H(X_i)-H(X_j)&=H(\boldsymbol o_i-\boldsymbol o_i'\,|\,X_i,X_j)+H(\boldsymbol o_j\,|\,X_i,X_j)\\
&\leq H(\alpha\boldsymbol o_i+\beta\boldsymbol o_j\,|\,X_i,X_j)+H(\alpha\boldsymbol o_i'+\beta\boldsymbol o_j\,|\,X_i,X_j)\\
&=2H(\by\,|\,X_i,X_j)\\
&\leq 2H(X_k)+2\delta_L,
\end{align*}
that is, $g_i\leq \sigma_i+\sigma_j$; exchanging $j$ and $k$ gives $g_i\leq \sigma_i+\sigma_k$. The first equality uses~\eqref{eq:cell_markov}, \eqref{eq:cell_bin} and that $\boldsymbol o_i'$ is generated from $X_i$ and fresh randomness. The inequality is~\eqref{eq:triangle} with $\bxi_1=\alpha\boldsymbol o_i$, $\bxi_2=\alpha\boldsymbol o_i'$ and $\bxi_3=-\beta\boldsymbol o_j$, where $\bxi_3$ is independent of $(\bxi_1,\bxi_2)$ given $(X_i,X_j)$ by~\eqref{eq:cell_markov} and the construction of $\boldsymbol o_i'$. The second equality holds because $(\boldsymbol o_i',\boldsymbol o_j)$ has the same conditional distribution as $(\boldsymbol o_i,\boldsymbol o_j)$, and the last step is~\eqref{eq:cell_dem}.
\end{proof}

\subsection{Proof of Theorem~\ref{thm:ve}}

\emph{Part 1: cross-ratios outside the prime subfield.} For distinct $i,j,k$, the demand is uniform and independent of $\boldsymbol o_k$, and $X_k$ is a function of $\boldsymbol o_k$, so
\begin{equation*}
L=H(\by\,|\,\boldsymbol o_k)\leq H(X_i,X_j\,|\,\boldsymbol o_k)+H(\by\,|\,X,\boldsymbol o_k)\leq L(R_i+R_j)+\delta_L.
\end{equation*}
These cut-set bounds, as $L\to\infty$, and the zero-error scheme of Theorem~\ref{thm:additive} give $\mathcal C^{\mathrm{ve}}\subseteq\cR_{\mathrm{cut}}=\mathcal C\subseteq\mathcal C^{\mathrm{ve}}$.

\emph{Part 2: cross-ratios in the prime subfield.} Let $\nu=\nu(\lambda)$ and $r=\nu-1$; since $2\leq\nu\leq(p+1)/2$, $1\leq r\leq(p-1)/2$, so $r$ is not a multiple of $p$. Relabel the users and scale so that $\lambda=\nu$, giving $\by=\ba+(r+1)\bb$. Apply Lemma~\ref{lem:entropy} conditionally on $(X_2,X_3)$, with $\bxi_1=\boldsymbol o_2$, $\bxi_2=-\boldsymbol o_3/r$, $\bxi_1'=\boldsymbol o_2'$ and $\bxi_2'=-\boldsymbol o_3'/r$, and average over the messages; given $(X_2,X_3)$, $\boldsymbol o_2$ and $\boldsymbol o_3$ are independent by~\eqref{eq:cell_markov}. Then
\begin{equation*}
\bxi_1+r\bxi_2=-\boldsymbol o_1,\qquad\bxi_1-\bxi_2=\by/r,
\end{equation*}
and by~\eqref{eq:cell_markov} and~\eqref{eq:cell_bin} the conditional entropy of $\bxi_1$ is $L-H(X_2)$ and the averages of the conditional growth terms are $g_2$ and $g_3$; moreover, the average of their minimum is at most $\min\{g_2,g_3\}$. Hence~\eqref{eq:key} gives
\begin{equation}\label{eq:ve_step}
H(\boldsymbol o_1\,|\,X_2,X_3)-H(\by\,|\,X_2,X_3)\leq(r-1)\big[H(\by\,|\,X_2,X_3)-\big(L-H(X_2)\big)+g_2\big]+\min\{g_2,g_3\}.
\end{equation}
We bound the cost of the steps on the right by $\sigma$.

\emph{Each addition costs at most $\sigma$.} By~\eqref{eq:cell_dem} and~\eqref{eq:cell_diff},
\begin{equation*}
H(\by\,|\,X_2,X_3)-\big(L-H(X_2)\big)\leq \sigma_3,\qquad g_2\leq \sigma_2+\sigma_1,
\end{equation*}
so the bracket in~\eqref{eq:ve_step} is at most $\sigma$.

\emph{The reflection costs at most $2\sigma/3$.} By~\eqref{eq:cell_diff}, $g_2\leq \sigma_2+\min\{\sigma_1,\sigma_3\}$ and $g_3\leq \sigma_3+\min\{\sigma_1,\sigma_2\}$. If $\sigma_1$ is the largest slack, $g_2\leq \sigma_2+\sigma_3$; if $\sigma_2$ is, $g_3\leq \sigma_3+\sigma_1$; if $\sigma_3$ is, $g_2\leq \sigma_2+\sigma_1$. Hence
\begin{equation*}
\min\{g_2,g_3\}\leq \sigma-\max\{\sigma_1,\sigma_2,\sigma_3\}\leq\frac{2\sigma}{3},
\end{equation*}
and~\eqref{eq:ve_step} gives
\begin{equation*}
H(\boldsymbol o_1\,|\,X_2,X_3)-H(\by\,|\,X_2,X_3)\leq\Big(r-1+\frac23\Big)\sigma.
\end{equation*}

\emph{The cut-set bounds require a large difference.} By~\eqref{eq:cell_sum} and~\eqref{eq:cell_dem},
\begin{equation*}
H(\boldsymbol o_1\,|\,X_2,X_3)-H(\by\,|\,X_2,X_3)\geq 2L-2\delta_L-\big(H(X_1)+H(X_2)+H(X_3)\big)=\frac{L-\delta_L-\sigma}{2}.
\end{equation*}

Combining the last two displays, $(L-\delta_L-\sigma)/2\leq(r-1/3)\sigma$, that is, $\sigma\geq 3(L-\delta_L)/(6r+1)$. Since $H(X_1)+H(X_2)+H(X_3)=\big(3(L-\delta_L)+\sigma\big)/2$ and $H(X_i)\leq LR_i$,
\begin{equation*}
R_1+R_2+R_3\geq\Big(\frac32+\frac{3}{2(6r+1)}\Big)\Big(1-\frac{\delta_L}{L}\Big).
\end{equation*}
Letting $L\to\infty$ and substituting $r=\nu-1$ proves the first bound in~\eqref{eq:ve_prime}, since $3/2+3/(2(6\nu-5))=3(3\nu-2)/(6\nu-5)$. The expression decreases with $\nu$, so $\nu\leq(p+1)/2$ gives the second, and the gap above $3/2$ shows $\mathcal C^{\mathrm{ve}}\neq\cR_{\mathrm{cut}}$. For harmonic $\lambda$, $\nu=2$ gives $12/7$, and two users sending their observations achieve sum-rate $2$.

\emph{Part 3: linear and additive encoders.} The argument applies to the general problem. Consider $\Fq$-linear or additive encoders, with the data and demand written in $\Fp$-coordinates in the latter case, and stack the messages as $\mathbf G\bw^L$, as in Footnote~\ref{fn:linear-decoder}. Given the messages, $\bw^L$ is uniform on a shift of $\nullspace(\mathbf G)$, so the demand $\mathbf D^{[L]}\bw^L$ is uniform on a shift of the subspace $\mathbf D^{[L]}\,\nullspace(\mathbf G)$, which does not depend on the messages. If this subspace is nonzero, it has at least $p\geq2$ elements, so every decoder is correct with probability at most $1/2$ given each value of the messages, and $\epsilon_L\geq 1/2$. Hence every code with $\epsilon_L<1/2$ has $\mathbf D^{[L]}\bw^L=\boldsymbol 0$ whenever $\mathbf G\bw^L=\boldsymbol 0$, and by Footnote~\ref{fn:linear-decoder} a linear decoder recovers the demand with zero error from the same messages. Thus $\mathcal C^{\mathrm{ve}}_{\mathrm{lin}}=\mathcal C_{\mathrm{lin}}$ and $\mathcal C^{\mathrm{ve}}_{\mathrm{add}}=\mathcal C_{\mathrm{add}}$ for the general problem; for the four-line instance, Theorem~\ref{thm:linear} and Corollary~\ref{cor:additive} identify these regions.

\begin{proof}[Proof of Proposition~\ref{prop:ve_twouser}]
The converse in the proof of Theorem~\ref{thm:twouser} (Appendix~\ref{app:twouser}) is entropic. With $\delta_L=H(\by\,|\,X_1,X_2)$ and the notation of that proof, $\kappa_1L=H(\by\,|\,\boldsymbol u_0,\boldsymbol u_2)\leq H(X_1\,|\,\boldsymbol u_0)+\delta_L$, since $X_2$ is a function of $(\boldsymbol u_0,\boldsymbol u_2)$, and in the same way $H(X_2\,|\,\boldsymbol u_0)\geq\kappa_2L-\delta_L$. The block $\by_0$ is a function of $\boldsymbol u_0$ and of $\by$, so $I(X_1,X_2;\boldsymbol u_0)\geq I(X_1,X_2;\by_0)\geq\kappa_{12}L-\delta_L$. As in that proof, $\log_q|\cX_1||\cX_2|\geq(\kappa_1+\kappa_2+\kappa_{12})L-3\delta_L$, and Fano's inequality, with a demand of $\dim\mathcal D$ symbols per instance, gives $\delta_L/L\to 0$. The zero-error codes of Theorem~\ref{thm:twouser} achieve the region.
\end{proof}

\subsection{What does not carry over}

Two aspects of the zero-error results remain unresolved under vanishing error. First, the exact harmonic capacity: the vanishing-error capacity for harmonic $\lambda$ lies in $[12/7,2]$; the message triple need not determine $(\ba,\bb)$ when the decoding error probability is positive, and reconstruction up to $o(L)$ uncertainty remains open. Second, the weighted zero-error bounds~\eqref{eq:peruser} have not been established under vanishing error; Theorem~\ref{thm:ve} gives different bounds. The unresolved harmonic case leaves the characterization in characteristics three and five (Corollary~\ref{cor:complete}) incomplete, since every $\lambda\in\Fp\setminus\{0,1\}$ is harmonic for $p\in\{3,5\}$; characteristic two carries over completely, since $\F_2\setminus\{0,1\}$ is empty. Finally, the results use uniform data: under vanishing error a non-uniform distribution can concentrate on a different set of inputs, and non-uniform data are not treated here.

\section{Relation to repair}\label{app:repair}

This appendix interprets the four-line results as statements about exact repair of scalar linear $[4,2]$ MDS storage codes. The repair schemes may be arbitrary, including nonlinear schemes. We describe the correspondence and compare the results with the repair literature.

A scalar linear $[4,2]$ MDS code over $\Fq$ stores one symbol on each of four nodes per codeword. If one node fails, each of the other three nodes sends a message to a replacement, which must recover the failed symbol with zero error. A repair scheme may operate jointly on $L$ independent codewords. Taking the helpers as users and the failed symbol as the demand gives the four-line instance~\eqref{eq:code}, after a change of basis and scaling as in Section~\ref{sec:fourline_model}.

In the language of the repair literature, $\mathcal Q_\lambda$ is a Reed--Solomon code: its codewords are the evaluations of the polynomials $a+bt$ in the variable $t$, of degree less than $2$, at $t=0,\infty,1,\lambda$, where evaluation at $\infty$ gives the coefficient $b$. Thus its coordinates are $a,b,a+b,a+\lambda b$, and $\lambda$ is the cross-ratio of the evaluation points. For $q=3$, the four points are $0,\infty,1,2$; they correspond to the four distinct lines represented by $(1,0)$, $(0,1)$, $(1,1)$ and $(1,2)$ in $\F_3^2$.

Under this correspondence, $R_i$ is the bandwidth of helper $i$ in symbols per codeword, $\cR_{\mathrm{cut}}$ is the region given by the cut-set bound~\cite{Dimakis_etal} for every subset of helpers, and $\cR_{\mathrm{all}}$ is achieved by downloading two nodes in full. The block length $L$ is the sub-packetization over $\Fq$. An additive code is a repair scheme linear over the prime subfield, as in~\cite{Shanmugam_Papailiopoulos_Dimakis_Caire,Guruswami_Wootters}, with sub-packetization $nL$ over $\Fp$.

For a scalar linear $[4,2]$ code over $\F_{p^n}$ with cross-ratio $\lambda$, Theorem~\ref{thm:main} gives the following. The bandwidth region is the cut-set region if and only if $\lambda\notin\Fp$; it is then achieved by a scheme linear over $\Fp$ that repairs two codewords jointly. For harmonic $\lambda$ no repair scheme, linear or not and at any sub-packetization, beats downloading two nodes, since every scheme reveals all the data to the replacement. For every $\lambda\in\Fp$ no scheme approaches the cut-set bandwidth $3/2$ at any sub-packetization: the bandwidth is at least $3p/(2p-1)$ symbols per codeword. Over fields of characteristic $2$, $3$ and $5$ the bandwidth region of every scalar linear $[4,2]$ MDS code is determined. Table~\ref{tab:repair} separates these statements from what was known.

\begin{table}[H]
\centering
\small
\renewcommand{\arraystretch}{1.25}
\begin{tabular}{p{4.6cm}p{4.9cm}p{4.9cm}}
\toprule
Repair of one node of a scalar linear $[4,2]$ MDS code over $\F_{p^n}$ & Known & This paper\\
\midrule
Cut-set bound & bandwidth $\geq 3/2$~\cite{Dimakis_etal} & tight if and only if $\lambda\notin\Fp$ (Theorem~\ref{thm:main})\\
\hdashline
$\lambda$ outside a subfield & achieved by repair linear over $\F_{p^{n/2}}$ when $n$ is even and $\lambda\notin\F_{p^{n/2}}$~\cite{Shanmugam_Papailiopoulos_Dimakis_Caire} & achieved for every $n$ when $\lambda\notin\Fp$, linear over $\Fp$, two codewords (Theorem~\ref{thm:additive})\\
\hdashline
Prime fields, nonlinear & approaches $3/2$ for almost all $\lambda$ as $p\to\infty$, one codeword~\cite{Con_Tamo} & additive schemes need $2$ (Corollary~\ref{cor:additive}); no scheme approaches $3/2$ for fixed $p$ (Theorem~\ref{thm:plunnecke})\\
\hdashline
Harmonic $\lambda$ & not known & no gain over two downloads, any scheme, any sub-packetization (Theorem~\ref{thm:harmonic})\\
\hdashline
$\lambda\in\Fp$ & lower bounds no stronger than the cut-set bound; symmetric schemes on one codeword~\cite{Con_Tamo} & at least $3p/(2p-1)$, any scheme, any sub-packetization (Theorem~\ref{thm:plunnecke})\\
\hdashline
Complete region & not known & every $\lambda$ for $p\in\{2,3,5\}$ (Corollary~\ref{cor:complete})\\
\bottomrule
\end{tabular}
\caption{Repair of scalar linear $[4,2]$ MDS codes.}\label{tab:repair}
\end{table}

Table~\ref{tab:repair} focuses on repair of scalar linear $[4,2]$ MDS codes. The broader literature on repair of Reed--Solomon codes, including~\cite{Guruswami_Wootters,Tamo_Ye_Barg,Duursma_Dau,Fang_Qiu_Xia_Fu,Li_cutset}, studies longer codes, mostly with linear repair schemes. Here, the converses of Theorems~\ref{thm:harmonic} and~\ref{thm:plunnecke} apply to arbitrary repair schemes for $[4,2]$ codes, allowing nonlinear encoders, joint repair of any number of codewords and unequal helper bandwidths.

\section{Proofs of Theorem~\ref{thm:twouser} and Proposition~\ref{prop:threescalar}}\label{app:twouser}

We first prove the two-user capacity region, then use it to classify three users with one-dimensional observations.

\begin{proof}[Proof of Theorem~\ref{thm:twouser}]
\emph{Change of basis.} Choose a basis of $\mathcal H_1\cap\mathcal H_2$, extend it separately to bases of $\mathcal H_1$ and $\mathcal H_2$, and extend their union to a basis of $\Fq^\mu$. Do the same for the demand: choose a basis of $\mathcal D\cap\mathcal H_1\cap\mathcal H_2$, extend it separately to bases of $\mathcal D\cap\mathcal H_1$ and $\mathcal D\cap\mathcal H_2$, and extend their union to a basis of $\mathcal D$. By the invariance in Section~\ref{sec:invariance}, with $\mathbf P$, $\mathbf Q_k$ and $\mathbf Q$ given by these bases, we may assume that $\bw=(\boldsymbol u_0,\boldsymbol u_1,\boldsymbol u_2,\boldsymbol u_3)$, that users $1$ and $2$ observe $(\boldsymbol u_0,\boldsymbol u_1)$ and $(\boldsymbol u_0,\boldsymbol u_2)$, and that the demand is
\begin{equation*}
\by=\mathbf D\bw=(\by_0,\by_1,\by_2,\by_3),
\end{equation*}
where $\by_0$ is a function of $\boldsymbol u_0$, $\by_k$ is a function of $(\boldsymbol u_0,\boldsymbol u_k)$ for $k=1,2$, and $\by_3$, since $\mathcal D\subseteq\mathcal H_1+\mathcal H_2$, is the sum of a linear function of $(\boldsymbol u_0,\boldsymbol u_1)$ and a linear function of $\boldsymbol u_2$; no part of the demand involves $\boldsymbol u_3$. Counting dimensions, $\by_0,\by_1,\by_2,\by_3$ have $\kappa_{12}$, $\kappa_1-\kappa_0$, $\kappa_2-\kappa_0$ and $\kappa_0$ entries, with $\kappa_0$ as in Remark~\ref{rem:gap}.

\emph{Converse.} Fix a zero-error code of block length $L$; in the converse, all vectors are stacked over the $L$ instances. Given $(\boldsymbol u_0,\boldsymbol u_2)$, the entries of $(\by_1,\by_3)$ are, up to known terms, linearly independent functions of $\boldsymbol u_1$, since a nontrivial combination not involving $\boldsymbol u_1$ would be a demanded function in $\mathcal H_2$, hence a combination of the entries of $(\by_0,\by_2)$, contradicting that the entries of $\by$ are linearly independent. Since $X_2$ is a function of $(\boldsymbol u_0,\boldsymbol u_2)$, $\by$ is determined by $(X_1,\boldsymbol u_0,\boldsymbol u_2)$. Hence
\begin{equation*}
\kappa_1L=H(\by\,|\,\boldsymbol u_0,\boldsymbol u_2)
\leq H(X_1\,|\,\boldsymbol u_0,\boldsymbol u_2)
=H(X_1\,|\,\boldsymbol u_0).
\end{equation*}
Similarly, $H(X_2\,|\,\boldsymbol u_0)\geq\kappa_2L$. These inequalities give $R_1\geq\kappa_1$ and $R_2\geq\kappa_2$. Since $X_1$ and $X_2$ are independent given $\boldsymbol u_0$, and $\by_0$ is a function of $\boldsymbol u_0$ and, by zero error, of $(X_1,X_2)$,
\begin{align*}
L(R_1+R_2)&=\log_q\bigl(|\cX_1||\cX_2|\bigr)\geq H(X_1,X_2)\\
&=H(X_1\,|\,\boldsymbol u_0)+H(X_2\,|\,\boldsymbol u_0)+I(X_1,X_2;\boldsymbol u_0)\\
&\geq(\kappa_1+\kappa_2)L+H(\by_0)
=(\kappa_1+\kappa_2+\kappa_{12})L.
\end{align*}

\emph{Achievability.} With $L=1$, user $1$ sends $\by_0$, $\by_1$ and the part of $\by_3$ depending on $(\boldsymbol u_0,\boldsymbol u_1)$, and user $2$ sends $\by_2$ and the part of $\by_3$ depending on $\boldsymbol u_2$. The server adds the two parts to obtain $\by_3$. The rate tuple is $(\kappa_{12}+(\kappa_1-\kappa_0)+\kappa_0,\,(\kappa_2-\kappa_0)+\kappa_0)=(\kappa_1+\kappa_{12},\kappa_2)$; letting user $2$ send $\by_0$ instead gives $(\kappa_1,\kappa_2+\kappa_{12})$. Time sharing and upward closure give the region, with $\Fq$-linear codes.
\end{proof}

\begin{proof}[Proof of Proposition~\ref{prop:threescalar}]
\emph{Merging.} If $\boldsymbol v_i$ and $\boldsymbol v_j$ are proportional, the pair $(X_i,X_j)$ is a function of one symbol per instance, so a code for the three users is a code for two users, one observing $\boldsymbol v_i^\top\bw$ with rate $R_i+R_j$; conversely the $\Fq$-linear codes of Theorem~\ref{thm:twouser} for that user can be split into two parts of any rates adding to its rate, with arbitrary rate splits obtained by time sharing and closure. If all three $\boldsymbol v_k$ are proportional, the problem has one user.

\emph{Independent lines.} If the $\boldsymbol v_k$ are linearly independent, the users observe disjoint coordinates after a change of basis, each user sends the part of the demand that involves its coordinate, and this meets the cut-set bound.

\emph{Three lines in a plane.} Let the $\boldsymbol v_k$ be pairwise independent and span a plane, so that any two users determine the third. If $\dim\mathcal D=2$, the cut-set region is $\{R_i+R_j\geq 1,\ R_1+R_2+R_3\geq 2\}$, whose vertices, the permutations of $(1,1,0)$, are achieved by two users sending their observations. If $\mathcal D=\mathcal H_1$, say, the cut-set bounds are $R_1+R_2\geq 1$ and $R_1+R_3\geq 1$, the others being implied. The vertices of this region are $(1,0,0)$, achieved by user $1$ sending the demand, and $(0,1,1)$, achieved by users $2$ and $3$ sending their observations; time sharing and upward closure give the region. If $\mathcal D$ is a line in the plane different from the three users' lines, the configuration is four distinct lines, the four-line instance~\eqref{eq:code} by Section~\ref{sec:fourline_model}.
\end{proof}

\bibliography{refs}

\end{document}